\documentclass[11pt,english]{article}
\pdfoutput=1
\usepackage{jheppub}
\usepackage{lscape}
\usepackage{amsmath}
\usepackage{slashed}
\usepackage{mathtools}
\usepackage{tikz}
\usetikzlibrary{shapes,arrows,shadows}
\usepackage{amsmath,bm,times}

\def\p{\partial}

\def\-{\hphantom{-}}

\def\s2{\frac{1}{\sqrt2}}

\def\be{\begin{equation}}
\def\ee{\end{equation}}
\def\bea{\begin{align}}
\def\eea{\end{align}}
\def\beqa{\begin{eqnarray}}
\def\eeqa{\end{eqnarray}}

\def\mg{m_{3/2}}
\def\mg2{m^2_{3/2}}

\def\Dsl{\,\raise.15ex\hbox{/}\mkern-13.5mu D} 

\newcommand{\eq}[1]{\begin{equation}
                     \begin{split} #1 \end{split}
                     \end{equation}}
   
   \newcommand{\La}{ \begin{picture}(0,0) 
      \put(-20,130){(a)}
  \end{picture}}
   \newcommand{\Lb}{ \begin{picture}(0,0) 
      \put(-20,130){(b)}
  \end{picture}}
   \newcommand{\Lc}{ \begin{picture}(0,0) 
      \put(-20,130){(c)}
  \end{picture}}
   \newcommand{\Ld}{ \begin{picture}(0,0) 
      \put(-20,130){(d)}
  \end{picture}}
   \newcommand{\Le}{ \begin{picture}(0,0) 
      \put(-20,130){(e)}
  \end{picture}}
   \newcommand{\Lf}{ \begin{picture}(0,0) 
      \put(-20,130){(f)}
  \end{picture}}
     \newcommand{\Laa}{ \begin{picture}(0,0) 
      \put(-180,130){(a)}
  \end{picture}}
   \newcommand{\Lbb}{ \begin{picture}(0,0) 
      \put(-180,130){(b)}
  \end{picture}}
   \newcommand{\Lcc}{ \begin{picture}(0,0) 
      \put(-180,130){(c)}
  \end{picture}}

\begin{document}

\title{Testing Swampland Conjectures with Machine Learning}

\author[a,d]{Nana Cabo Bizet,} 
\author[b,d]{Cesar Damian,}
\author[a,d]{Oscar Loaiza-Brito$^3$,}
\author[c,d]{Dami\'an Kaloni Mayorga Pe\~na,}
\author[b,d]{J. A. Monta\~nez-Barrera}
\affiliation[a]{Departamento de F\'isica, Universidad de Guanajuato,
Loma del Bosque No. 103 Col. Lomas del Campestre C.P 37150 Leon, Guanajuato, Mexico.}
\affiliation[b]{Departamento de Ingenier\'ia Mec\'anica, Universidad de Guanajuato,
Carretera Salamanca-Valle de Santiago Km 3.5+1.8 Comunidad de Palo Blanco, Salamanca, Mexico}
\affiliation[c]{Mandelstam Institute for Theoretical Physics, School of Physics, NITheP, and CoE-MaSS, University of the Witwatersrand, Johannesburg, WITS 2050, South Africa}
\affiliation[d]{Data Laboratory, Universidad de Guanajuato,
Loma del Bosque No. 103 Col. Lomas del Campestre C.P 37150 Leon, Guanajuato, Mexico.}

\emailAdd{nana@fisica.ugto.mx}
\emailAdd{cesaredas@fisica.ugto.mx}
 \emailAdd{oloaiza@fisica.ugto.mx}
 \emailAdd{damian.mayorgapena@wits.ac.za}
 \emailAdd{ja.montanezbarrera@ugto.mx}

\date{\today}

\abstract{We consider Type IIB compactifications on an isotropic torus $T^6$ threaded by geometric and non geometric fluxes. For this particular setup we apply supervised machine learning techniques, namely an artificial neural network coupled to a genetic algorithm, in order to obtain 
 more than sixty thousand flux configurations yielding to a scalar potential with at least one critical point. We observe that both stable AdS vacua with large moduli masses and small vacuum energy as well as unstable dS vacua with small tachyonic mass and large energy are absent, in accordance to the Refined de Sitter Conjecture. Moreover, by  considering a hierarchy among fluxes, we observe that perturbative solutions with small values for the vacuum energy and moduli masses are favored, as well as scenarios in which the lightest modulus mass is much greater than the corresponding AdS vacuum scale. Finally we apply some results on Random Matrix Theory to conclude that the most probable mass spectrum derived from  this string setup is that satisfying the Refined de Sitter and AdS scale conjectures.}
\arxivnumber{}


\maketitle



\section{Introduction}
One of the main aims 
of string theory is the construction of realistic effective theories with a small cosmological constant $\Lambda$ within the perturbative regime. Motivated by the recent series of conjectures around the construction of de Sitter (dS) vacua and inflationary conditions \cite{Garg:2018reu,Garg:2018zdg,Obied:2018sgi, Agrawal:2018own, Ooguri:2018wrx, Danielsson:2018qpa, Danielsson:2018ztv, Blumenhagen:2019qcg} (see also \cite{Damian:2013dq, Damian:2013dwa,CaboBizet:2016qsa, Bizet:2016paj,  Damian:2018tlf, Bedroya:2019snp, Andriot:2020lea, Lust:2019zwm, Palti:2019pca}), the question about a possible microscopic origin of $\Lambda$ has lately received an increasing attention \cite{Blumenhagen:2019vgj}. It is then worthwhile to focus on specific flux configurations  which can be related to effective models with small energy values at extremal points in moduli space. \\

In this context, one would be tracing back the origin of a small $\Lambda$ to some well-identified features of flux configurations. This would certainly be very interesting since fluxes drive many important physical phenomena, such as: supersymmetry breakdown, symmetry breaking, axion monodromy inflation and F-term monodromies. As it was observed in 
  \cite{Blumenhagen:2014gta, Blumenhagen:2015xpa, Blumenhagen:2015kja, Blumenhagen:2017cxt}, all these expected and desirable features naturally arise 
in the so called flux-scaling scenarios, where fluxes play a role
in fixing the values of the vacuum energy at extrema of the potential.
\\

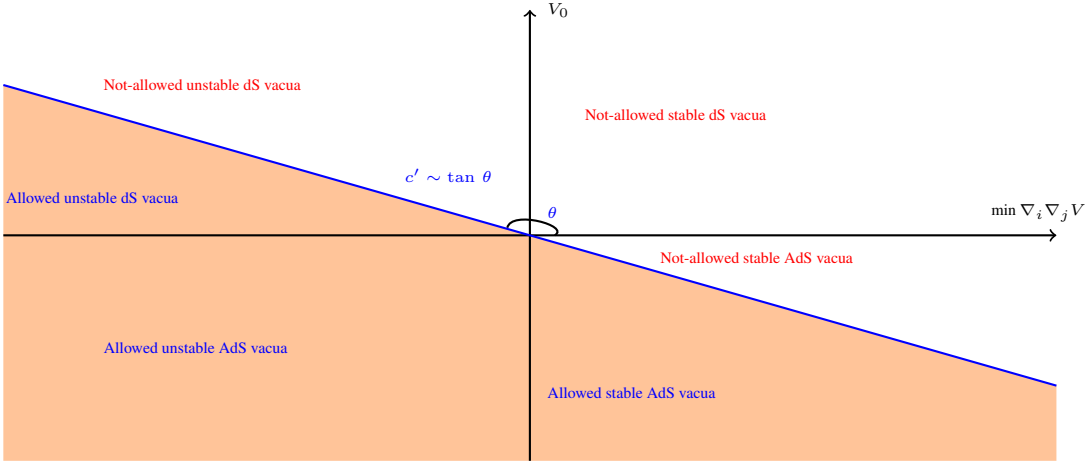
\begin{figure}[!htb]
\centering
\begin{tikzpicture}
\path[fill=white] (-7,2) -- (7,-2) -- (7,3) -- (-7,3) -- (-7,2);
\path[fill=orange!30!yellow!20!pink] (-7,2) -- (7,-2) -- (7,-3) -- (-7,-3) -- (-7,2);
\draw [thick] (-.3,.1) to [out=78,in=30] (.33,0);
\draw [black,thick,->] (-7,0) -- (0,0) -- (7,0);
\draw [black,thick,->] (0,-3) -- (0,0) -- (0,3);
\draw [blue,thick] (-7,2) -- (0,0) -- (7,-2);
\node at (-1.8,0.8) [right,blue] {\tiny $c' \sim \tan \, \theta$};
\node at (0.1,0.3) [right,blue] {\tiny $\theta$};
\node at (0.1,-2.1) [right,blue] {\tiny Allowed stable AdS vacua};
\node at (6.0,0.3) [right,black] {\tiny $\text{min}\, \nabla_i \nabla_j V$};
\node at (0.1,3.0) [right,black] {\tiny $V_0$};
\node at (1.6,-0.3) [right,red] {\tiny Not-allowed stable AdS vacua};
\node at (0.6,1.6) [right,red] {\tiny Not-allowed stable dS vacua};
\node at (-5.8,2.0) [right,red] {\tiny Not-allowed unstable dS vacua};
\node at (-5.8,-1.5) [right,blue] {\tiny Allowed unstable AdS vacua};
\node at (-7.1,0.5) [right,blue] {\tiny Allowed unstable dS vacua};
\end{tikzpicture}
\caption{\emph{Agreement of  given effective model with the refined de Sitter conjecture defines 6 different zones. Three of them belong to the Swampland. The angle $\theta$ defines the slope of the line dividing the regions of the Swampland from the Landscape.}}
\label{fig:RdSC}
\end{figure}

A promising scenario as they are, flux compactifications must obey the quantum gravity conjectures if one hopes to complete these models in the UV regime. In this work we focus on the so-called Refined de Sitter Conjecture (RdSC) which states that the construction of a stable dS vacuum is excluded from a consistent quantum gravity theory (including string compactifications). More specifically the RdSC  establishes a bound of the form 
\begin{equation}
\frac{\text{min} \,\nabla_i \nabla_j V }{V} \leq -c' \,,
\end{equation}
where $V$ is a given effective scalar potential and $i$, $j$ represent index coordinates in field space and $c'$ is a given constant parameter. Besides the exclusion of stable de Sitter, the bound also implies that some apparently plausible AdS vacua must be discarded as well, depending on the actual value of the constant $c'$ as shown in Figure \ref{fig:RdSC}. The bound defines a line with a slope determining the value of $c'$ for some specific model, i.e., the upper bound on the quotient between the minimum mass squared and the value of $V$ at an extremum for the potential. In Figure  \ref{fig:RdSC} we can distinguish six different zones depending on whether the corresponding vacuum energy is positive or negative and on whether the vacuum is stable or not. 
As it was mentioned already, some AdS regions are excluded as well, in particular, stable AdS vacua with small energy and large moduli masses. The same is true for unstable dS regions with a 
large vacuum energy and a small tachyonic mass.\\ 

%
We concentrate on a simple well studied model consisting on a Type IIB compactification on an isotropic torus in presence of orientifold 3-planes, threaded by the usual Ramond-Ramond (RR)  and Neveu-Schwarz$-$Neveu-Schwarz (NS-NS) 3-form fluxes and by non-geometric (nG) fluxes as well \cite{Cribiori:2019hrb, Betzler:2019kon, Blaback:2015zra, Blaback:2013ht, Blumenhagen:2013aia, Blumenhagen:2013hva, Plauschinn:2018wbo}
 (see Appendix A). The scalar potential has three complex scalar fields: the complex structure ($U$), the axio-dilaton ($S$) and the K\"ahler modulus ($T$). The simplicity of this model lets us implement an algorithm to find as many extrema as possible for the scalar potential. One of the goals of the present work is to produce consistent and adequately quantized flux configurations. This, in order to obtain a reasonable sample of scenarios where one would be able to test 
whether or not the stable AdS and non-stable dS zones are excluded, in accordance or disagreement with the RdS conjecture.  \\

We classify different flux configurations according to the features of the scalar potential at the extremum under consideration. 
For that purpose we use an Artificial Neural Network\footnote{Implementation of different computational methods in high-energy physics research has increased notably in the last few years. 
 See e.g. \cite {Ruehle:2017mzq, Carifio:2017bov, He:2018jtw, Erbin:2018csv, Mutter:2018sra, Ashmore:2019wzb, Parr:2019bta, Halverson:2019tkf, Cole:2019enn, Ruehle:2020jrk, Gal:2020dyc}} (ANN), by means of which we are able to classify more than sixty thousand different flux configurations and some relevant features of the corresponding vacua. There is however an important caveat here. It is necessary to provide the ANN with concrete examples for the network to be able to identify certain patterns among the different fluxes, which in turn would lead  to some stable or unstable extremal point in moduli space. This is the reason to use genetic algorithms previous to adapting the neural network \cite{Damian:2013dq, Damian:2013dwa,Abel:2014xta,Ruehle:2017mzq,Cole:2019enn,AbdusSalam:2020ywo}. Since there is not a single example of a stable dS, it is possible that the network does not identify such cases and in consequence it will not learn how to construct them. So, we expect not to find dS stable extrema. Observe that this fact is only a consequence of our algorithm and it is not reflecting a general feature of our compactification model. However, we are not restricting the possible AdS vacua to encounter since there are plenty of examples of unstable and stable AdS extrema. By looking for them employing the neural network, we expect to reproduce all possible situations. Therefore, this is a fruitful zone to check for consistency with the RdS conjecture, and we find that those zones excluded by it are indeed absent in our classification, suggesting the validity of the conjecture or the quantum gravitational consistency of the considered setup.\\


Based on recent results \cite{Damian:2018tlf, CaboBizet:2019sku, Betzler:2019kon}, in which the presence of hierarchical values on fluxes induces a natural hierarchy on moduli masses, for which there are concrete (supersymmetric and non-supersymmetric) vacuum solutions with a small value for the cosmological constant $\Lambda$, we contemplate the possibility that hierarchical flux configurations lead to scenarios with small values for the vacuum energy. 
We observe that indeed, the values of the scalar potential at its minimum are smaller than one 
when the flux configuration possesses a hierarchy among their integer values. In this sense we suggest that a possible microscopic explanation for a small $\Lambda$ in a quantum gravity theory such as string theory, might rely on specific features of the flux configuration. 
Moreover, we find that the smaller the string coupling, the higher the probability to find a vacuum solution with a small vaccum energy, suggesting that for the most probable scenarios, $\Lambda\sim \exp(-Re(S))$. This is another highlight of the use of hierarchical flux configurations.\\ 

We also report that, by considering hierarchical fluxes, the ANN classification shows that there is a higher probability for the vacuum solutions to show a spectrum in which the 
minimal stable modulus mass is greater than the scale of the AdS vacuum. These vacua, in accordance to the AdS scale conjecture 
cannot be uplifted to a stable dS vacuum. \\

In order to sustain the above observations on a more solid basis, 
we compare the spectra of critical values obtained from the mass matrix, with the spectra of a Gaussian Orthogonal Ensemble (GOE) with a mean-value $\mu$ and standard deviation $\sigma$. We observe that the mass matrix posseses similar characteristics as a GOE namely, the probability for the mass matrix eigenvalues to be non-negative coincides with that derived from  a GOE. Thus, we use the spectral results obtained from Random Matrix Theory applied to the squared mass eigenvalues to find that:
\begin{itemize}
\item Probability to find an unstable critical point is $10^{6}$ times higher than finding a stable one.
\item  $80\%$ of  all generated flux configuration fulfilling string constraints as Tadpole cancelation and Bianchi identities do not exhibit a hierarchy among their values, pointing out the fact that it is not  likely to obtain a hierarchical flux configuration from random selection.
\end{itemize}
Although the last point suggests that it is very unlikely to encounter 
a flux configuration with a hierarchy, 
 if one departs from a hierarchical flux configuration, the probability to obtain an effective theory at the extremum of the scalar potential with some desired physical properties increases. This is:
%
\begin{itemize}
\item $70\%$ of the constructed vacua are within the perturbative regime. 
\item Among all vacua (stable critical points), $40\%$ of them have an (absolute) energy value smaller than unit.   
\item In $80\%$ of all AdS vacua, the lightest moduli mass is larger than the (absolute value of) vacuum scale.   
\end{itemize}
Therefore, although all generated vacua seems to satisfy the RdS conjecture we find that by restricting the construction of these simple models to hierarchical flux configurations, we increase the probability for the effective models to be in the perturbative regime and tu fulfill the AdS scale conjecture as well. This suggests, at least for these simple toroidal models, that the source of the Swampland constraints could rely on specific features on flux configurations as the hierarchical values among them. \\

Our work is organized as follows: In section \ref{sec:dos} we describe generically and in simple terms the implementation of the Artificial Neural Network coupled to the Genetic Algorithm. Technical issues concerning the structure of an ANN as well as a basic example are given in Appendix \ref{sec:app3}. In section \ref{sec:tres} we discuss the numerical results obtained by implementing the scan over random fluxes and on hierarchical flux configurations. 
Finally in Section \ref{sec:cuatro} we present our concluding remarks. The physical description of the Type IIB flux compactification setup is presented in Appendix \ref{sec:app:torus}. Similarly, a toy example illustrating the 
possibility to have small vacuum energy values at an extremum of the scalar potential 
is presented in Appendix \ref{sec:app:torus} as well.

\section{Classification of Vacua and Search for Extrema of the Potential}
\label{sec:dos}

We are interested in classifying vacua constructed from different flux configurations. This is done in order to identify flux patterns which could lead to some desired particularities, such as: stability, a small value for the cosmological constant or the existence of a dS critical point. For that we shall use and implement an Artificial Neural Network (ANN)\footnote{This section deals with some technical details and numerical analysis (see Appendix \ref{sec:app3} for details). For the reader interested in our conclusions on the construction of string derived effective models we suggest to go directly to Section 3.}.\\

The ANN architecture proposed in this paper is that of a pattern recognition feedforward network organized in three clusters of neurons: The input layer with 10 neurons, the hidden layer consisting of 12 neurons for the case of free-tachyon classification and 23 for the case of positive vacua classification, and the output with 1 neuron. The activation functions are chosen to be the hyperbolic tangent sigmoid transfer function. In the input we encode the integral values for the flux parameters consisting on a set of fluxes satisfying all string constraints, namely the Tadpole cancellation condition and  Bianchi identities. In our case we consider \textit{non-geometric fluxes} as well. \\

As previously mentioned, we concentrate on an isotropic toroidal flux compactification (see Appendix \ref{sec:app:torus}). Hence we consider 4 integers parameterizing the R-R sector fluxes $f$ (with components $f_i, \, i=1...4$), 4 integers for  the NS-NS sector fluxes as well $h$ (with components $h_j, j=1...4$), and 6 for the non-Geometric (nG) fluxes $b$ (with components $b_k, k=1...6$), adding up to the 14 nodes of the input. The output is made of those vacuum solutions of the scalar potential constructed from the corresponding flux compactification. Extra criteria must be added to stimulate the ANN searching. In our case we shall analyze two different criteria to stimulate the ANN, namely by looking for stable or dS critical points.\\

The use of the ANN requires a controlled training as a first step. The training consists 
on feeding the ANN with different flux configurations for which we know the existence of critical points as well as their corresponding features, such as vacuum stability and the value of the scalar potential at the critical point. The training data is obtained by 
randomly generating different flux configurations satisfying the Tadpole cancelation condition and Bianchi identities. 
We were able to generate about 40,000 different configurations using Mathematica codes. After that, we implemented a Genetic Algorithm (GA) in order to compute the moduli VEVs at which the scalar 
potential has a critical point, the corresponding scalar potential value at that point as well as its corresponding Hessian matrix (determining the stability). \\

The training process serves to optimize the network parameters (weights and biases) upon stepwise minimization of a certain objective function, which we have chosen to be the mean standard error (MSE, see Eq. \ref{eq:mse}). For this purpose, the training data is divided into three randomly selected groups as follows: 80$\%$ of the data is used for the ANN training, 10$\%$ for validation (to avoid overfitting on the training data), and 10 $\%$ for a posterior test (to avoid overfitting on the validation data)\footnote{\label{fn} 
Once the network is trained, the confusion matrix shows us that the ANN was able to correctly characterize 
the data in 98.6 $\%$ of the cases. For 75.2 $\%$ of the correct classification the output was positive (no tachyons in the spectra) and for 23.3 $\%$ of the correct classification the output was negative 
(there was at least one tachyon). Besides, the ANN made a wrong classification of the positive answer by 0.8 $\%$ (the ANN predicted at least one tachyon where there was no tachyon in the spectra) and it made a wrong classification with negative answer with an error of 0.6 $\%$ (the ANN predicted no tachyons where there was at least one tachyon in the spectrum).}. Thus it is expected from it to perform well beyond the training data (it might even be 
able to identify possible patterns relating the flux configuration with the existence of specific extrema of the resulting potential as well as the features of the potential at those critical points). 
The training data is divided into three randomly selected groups as follows: 80$\%$ of the data was used for the ANN training, 10$\%$ for the validation (to avoid overfitting on the training data), and 10 $\%$ for a posterior test (to avoid overfitting on the validation data). 

Once the ANN is trained we proceed to feed it with a variety of flux configurations. 
The ANN tells us which of them allow or not for the existence of some critical point with some required feature, i.e., it classifies the flux configurations into two groups according whether they fulfill the selected criteria or not. We confirm the results given by the ANN by implementing the GA and calculating specific values at the critical points in case we have them. A flow map of our approach is shown in 
Figure \ref{eq:arch}. More in detail, the sketch of our procedure is as follows:

\begin{enumerate}
\item We collect the training data. These are flux configurations fulfilling Tadpole cancellation condition and Bianchi identities. We generate nearly 40,000 different configurations. There are two training processes depending on the type of training data:
\begin{enumerate}
\item
\textbf{CASE A: Training with Random Fluxes}. The NS-NS, RR and nG fluxes are picked at random. 
\item
\textbf{CASE B: Training with Hierarchical Fluxes}. 
Fluxes used for training are no longer chosen at random. Instead, the flux values in one of the closed sectors are higher than the rest, e.g. integer valued NS-NS fluxes are between one and four orders of magnitude 
larger than R-R and nG fluxes.
\end{enumerate}

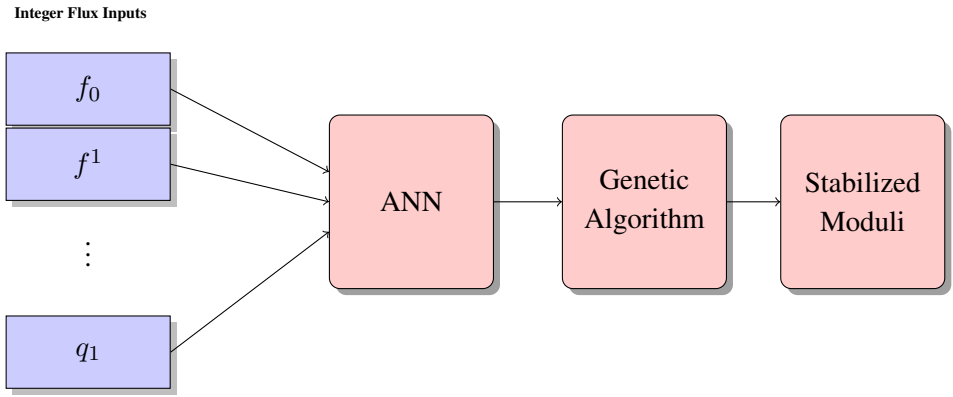
\begin{figure}[h!]
\centering
\tikzstyle{sensor}=[draw, fill=blue!20, text width=5em, 
    text centered, minimum height=2.5em,drop shadow]
\tikzstyle{ann} = [above, text width=5em, text centered]
\tikzstyle{wa} = [sensor, text width=5em, fill=red!20, 
    minimum height=6em, rounded corners, drop shadow]
\tikzstyle{sc} = [sensor, text width=13em, fill=red!20, 
    minimum height=10em, rounded corners, drop shadow]

\begin{tikzpicture}
    \node (wa) [wa]  {ANN};
    \node at (-4.4,2.5) {\bf{\tiny{Integer Flux Inputs}}};
    \path (wa.west)+(-3.2,1.5) node (asr1) [sensor] {$f_0$};
    \path (wa.west)+(-3.2,0.5) node (asr2)[sensor] {$f^1$};
    \path (wa.west)+(-3.2,-1.0) node (dots)[ann] {$\vdots$}; 
    \path (wa.west)+(-3.2,-2.0) node (asr3)[sensor] {$q_1$};    
   
    \path (wa.east)+(2.0,0) node (vote) [wa] {Genetic Algorithm};
   \path (vote.west)+(4.0,0) node (vote1) [wa] {Stabilized Moduli};

    \path [draw, ->] (asr1.east) -- node [above] {} 
        (wa.160) ;
    \path [draw, ->] (asr2.east) -- node [above] {} 
        (wa.180);
    \path [draw, ->] (asr3.east) -- node [above] {} 
        (wa.200);
    \path [draw, ->] (wa.east) -- node [above] {} 
        (vote.west);       
    \path [draw, ->] (vote.east) -- node [above] {} 
        (vote1.west);

\end{tikzpicture}
\caption{\emph{Flow chart of the vacua search procedure. One starts with a given flux configuration as an input for the neural network. The outcome is whether or not the fluxes under consideration lead to a scalar potential in the effective theory with critical points. If the outcome is positive, then one employs the Genetic Algorithm in order to find the critical point(s) and the corresponding field values at which the various moduli get fixed.} 
}
   \label{eq:arch}
\end{figure}

\item We use our trained network as a classifier for nearly 1.4 million flux configurations. 
In order to find some interesting statistics we have also selected two different criteria for the outcome data:
\begin{enumerate}
\item
\textbf{Criterion I: A stable critical point for the scalar potential}. This means that the ANN looks for patterns on the flux configuration such that the scalar potential has a minimum. This can be either AdS or dS 
\footnote{Minkowski vacua are excluded since, by construction the ANN is not trained to obtain such vacua.}.
\item
\textbf{Criterion II: A dS critical point}. This means that the network is asked to determine whether a given flux configuration exhibits a dS critical point, regardless of whether it is a maximum, a minimum or a saddle point. 
\end{enumerate}
\item We implement a Genetic Algorithm (GA) to compute specific values for the vacua on the classified flux configurations.
\end{enumerate}
In the following we describe our results by dividing them in terms of the flux configuration input set.\\

\subsection{Case A. Random Fluxes}

\subsection*{ANN Training}

After randomly generating 40 000 sets of fluxes satisfying the tadpoles and the Bianchi identities, we implement a GA to determine which of them contain critical points. We find 4034 critical points out of which there are 298 AdS solutions without tachyons, 139 dS with Tachyons and the remainder are tachyonic AdS. The results are used to train a network neural classification which assigns a value, e.g., 1 or 0 as an output, depending on whether or not a given property is satisfied by the flux under consideration.\\ 

As mentioned above we have selected two different cases according to the feature we want the ANN to find: 1) A stable critical point, this is, a minimum regardless the value of the vacuum energy or 2) A critical point with a positive value of the scalar potential at such point. This would be a dS critical point, regardless its stability.  For the first case, the ANN  classifies  flux configurations into 3 groups: Those generating a scalar potential with a stable critical point, those generating a scalar potential with unstable critical points and finally those generating a scalar potential without critical points. Similarly, for the second criterion, the classification of fluxes after feeding the ANN consists on a group of fluxes generating a dS extremal point, those with an AdS critical point, and finally, those generating a scalar potential without a critical point.

\subsection*{Results}

After training the ANN we feed it with nearly a million different flux configurations satisfying the Tadpole and Bianchi constraints. In the following we summarize our findings.\\ 

\begin{figure}[htbp]
   \centering
   \includegraphics[scale=0.28]{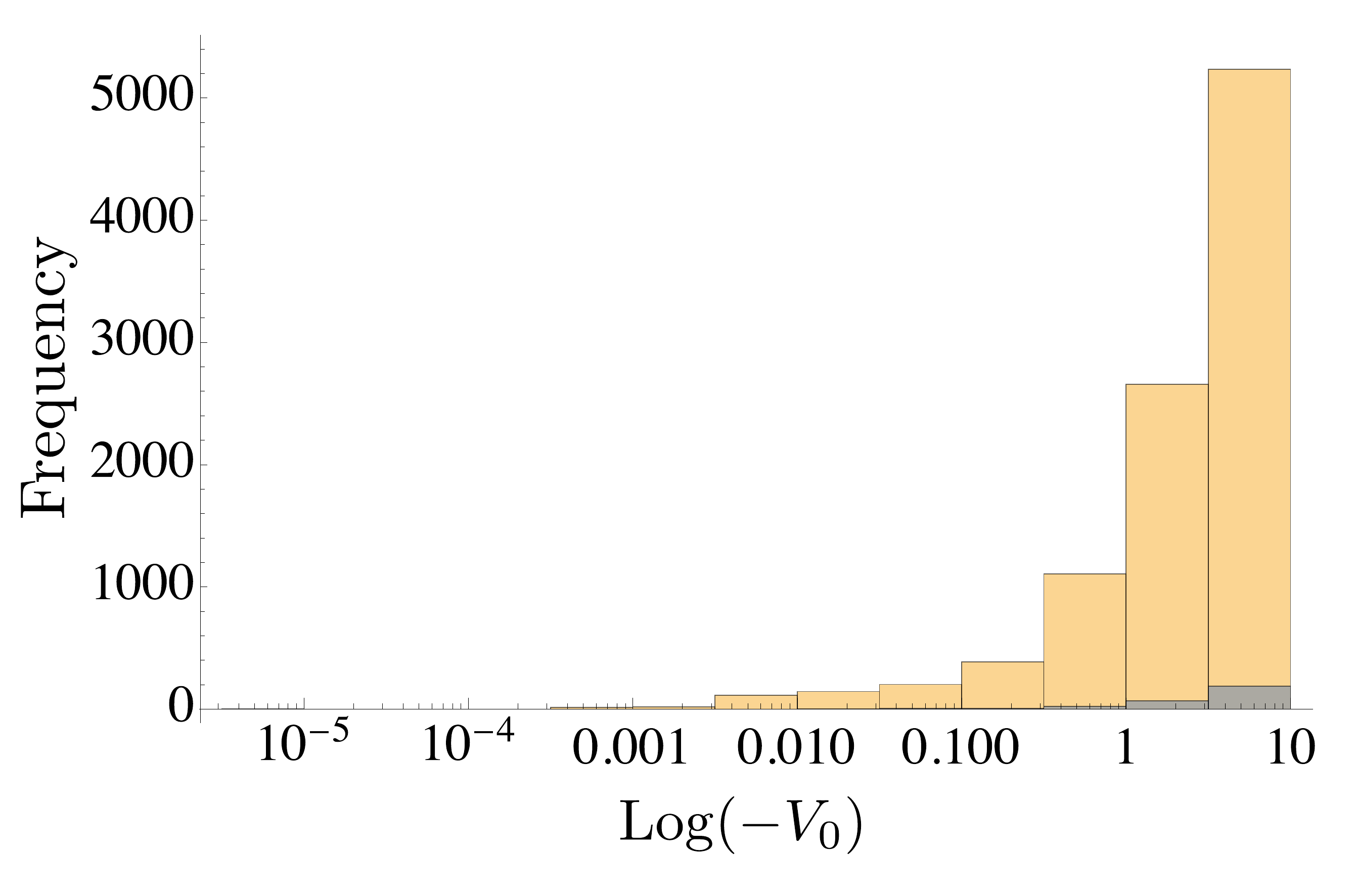}\,\,\, \includegraphics[scale=0.28]{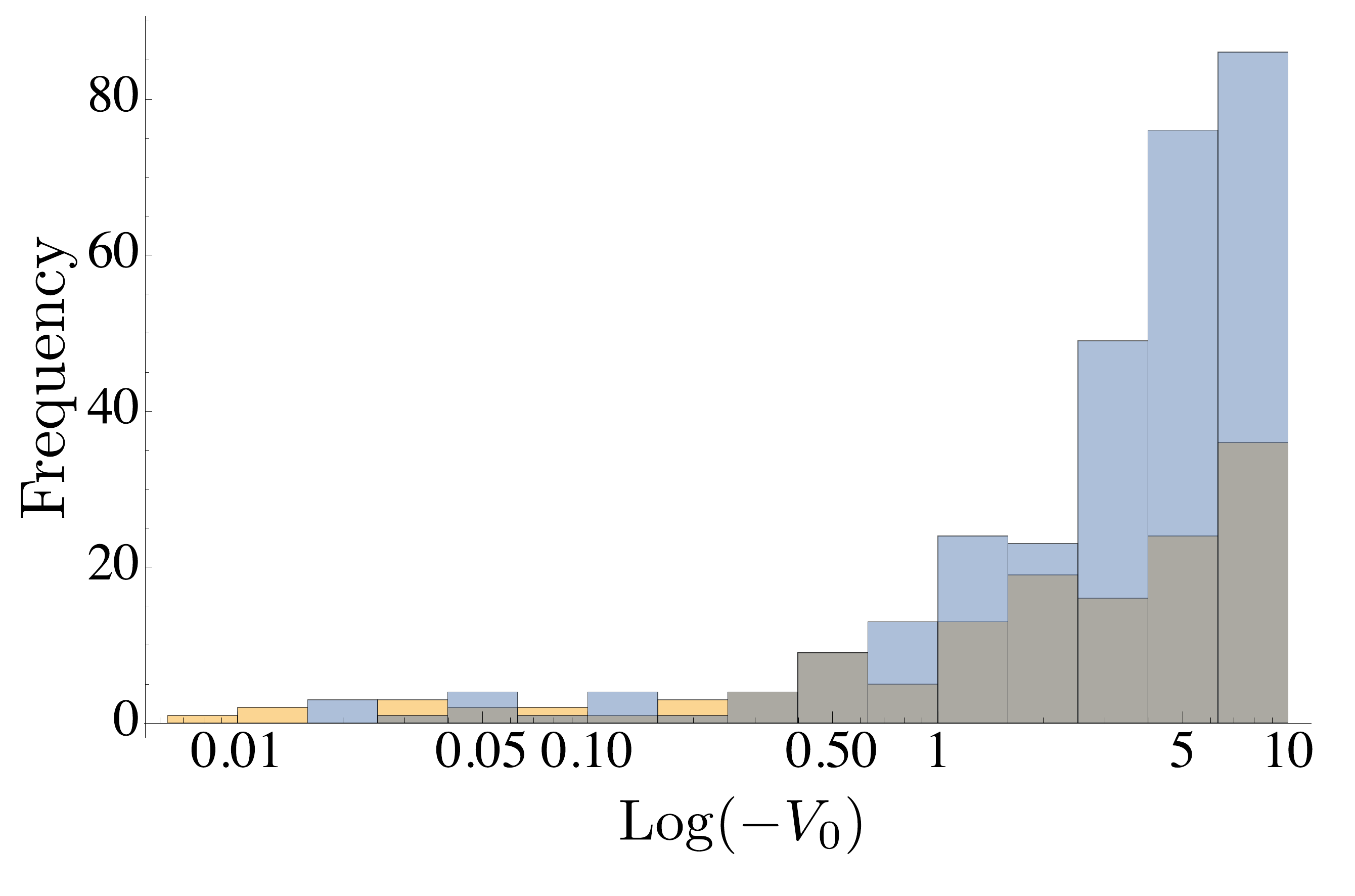}
   \caption{\emph{Histogram of the stable vacua with random fluxes generated by the GA (blue bars) and by the ANN+GA (yellow bars). Intersection of blue and yellow bars appears as gray bars. Left: Criterion I. Right: Criterion II. This data represents the first step in the training of the ANN and corresponds only to the free tachyon vacua.}}
      \label{fig:varvac}

\end{figure}

\noindent
\textbf{Criterion I. Stable Critical points}. 
Out of the roughly one million cases in the input, the ANN selects 66 000 sets of fluxes as candidates to generate a scalar potential with a minimum. In order to verify this, we use the GA and find that out of the 66 000 configurations, there are 20 779 with critical points and only 9872 without tachyons (see footnote \ref{fn}). 
It is interesting to compare with the original training data, out of 40.000 flux configurations we obtained  298 stable critical points, a naive estimate can lead us to the expectation of 7450 stable critical points had we simply run the AG over one million flux configurations. Employing the ANN coupled to the GA we obtain an amount of minima in the same order of magnitude (slightly higher). From this observation we conclude that besides the advantage of the ANN+GA being much less time consuming than the GA alone, we obtain roughly the same quality in the final outcome, therefore making this approach very suited for Landscape studies. The distribution of minima is presented in Figure \ref{fig:varvac} (a).  
Finally let us recall that no dS minimum was found, although there are many unstable dS extremal points.\\

\noindent
\textbf{Criterion II. dS extremal points.}
For this case the ANN 
favored  a total of 50 000 sets of fluxes as possible candidates to contain a dS extremum. The GA confirms that out of those 50 000, only 4944 different flux configurations generate a scalar potential with an extremal point.  Moreover, only 140 of them lead to a minimum, i.e., an extremal point free of tachyons.  For all of the stable minima we find that they occur at negative values of the scalar potential, i.e., they are AdS minima.
The rest of them correspond to unstable 2 744 dS and 2 200 AdS extremal points. The results of this classification are shown in Figure \ref{fig:varvac} (b)\footnote{Here we are analyzing only stable critical points out of those generated by the ANN through Criterion II.}.
Notice that contrary to training with Criterion I (9872 cases without tachyons), the number of stable vacua fund using Criterion II (140 cases without tachyons) is less than the one obtained by the use of GA on aleatory fluxes (298  cases without tachyons).\\

It is important to emphasize that from the total set of critical points, no stable dS vacua was found no matter what criterion we have used. For instance, with Criterion I, the number of dS (180 cases) is considerably smaller than those obtained in Criterion II (2744 cases). Also with Criterion II  the number of dS critical points increases as expected, in spite of an observed overall decrease in the number of stable points. This numerical analysis shows 
a correlation between the presence of tachyons and the number of actual dS critical points as suggested by the RdS Swampland Conjecture, at least for the isotropic torus with fluxes. \\

\begin{figure}[!htbp]
   \centering
\includegraphics[scale=0.28]{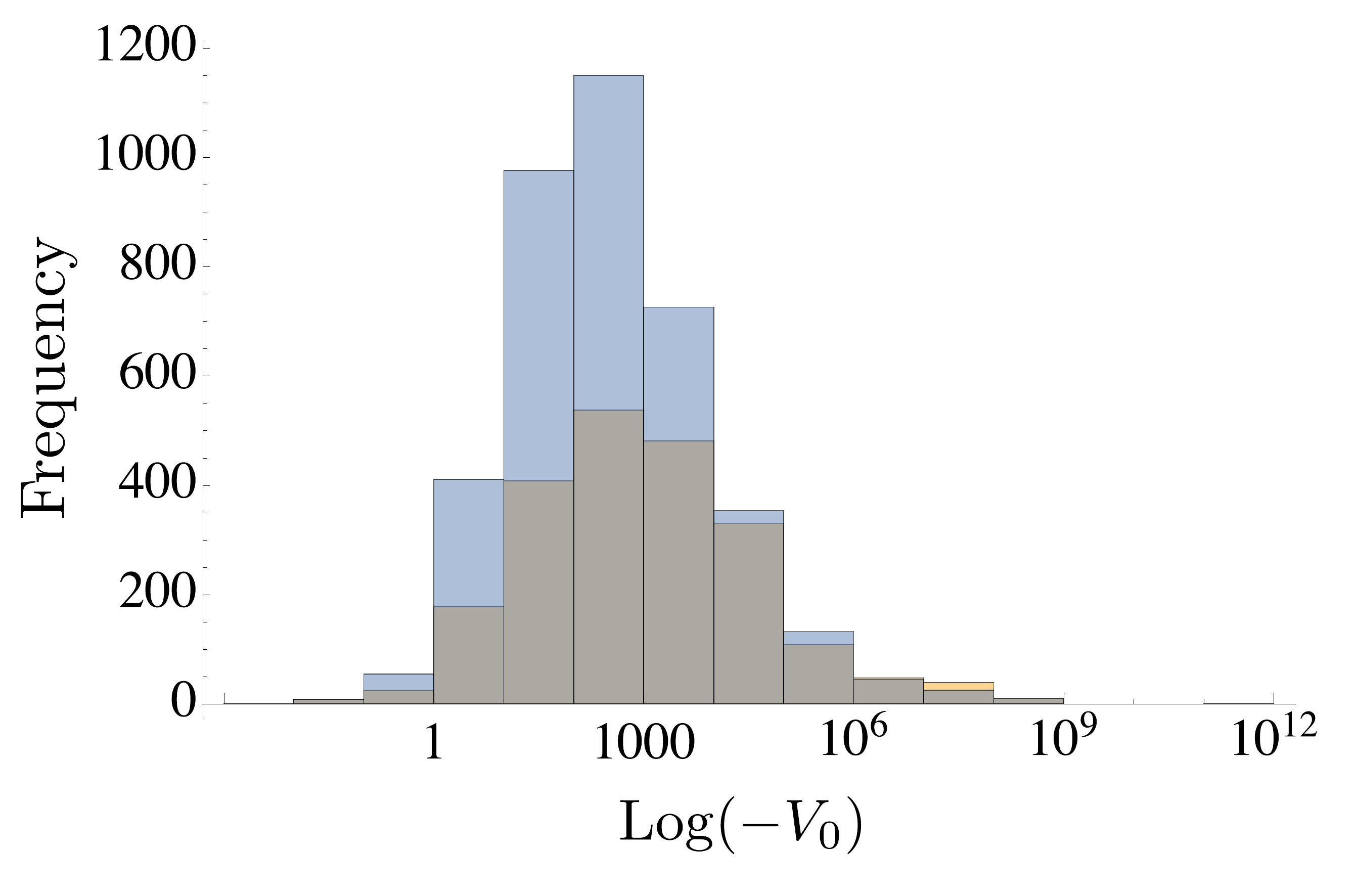}\La \includegraphics[scale=0.28]{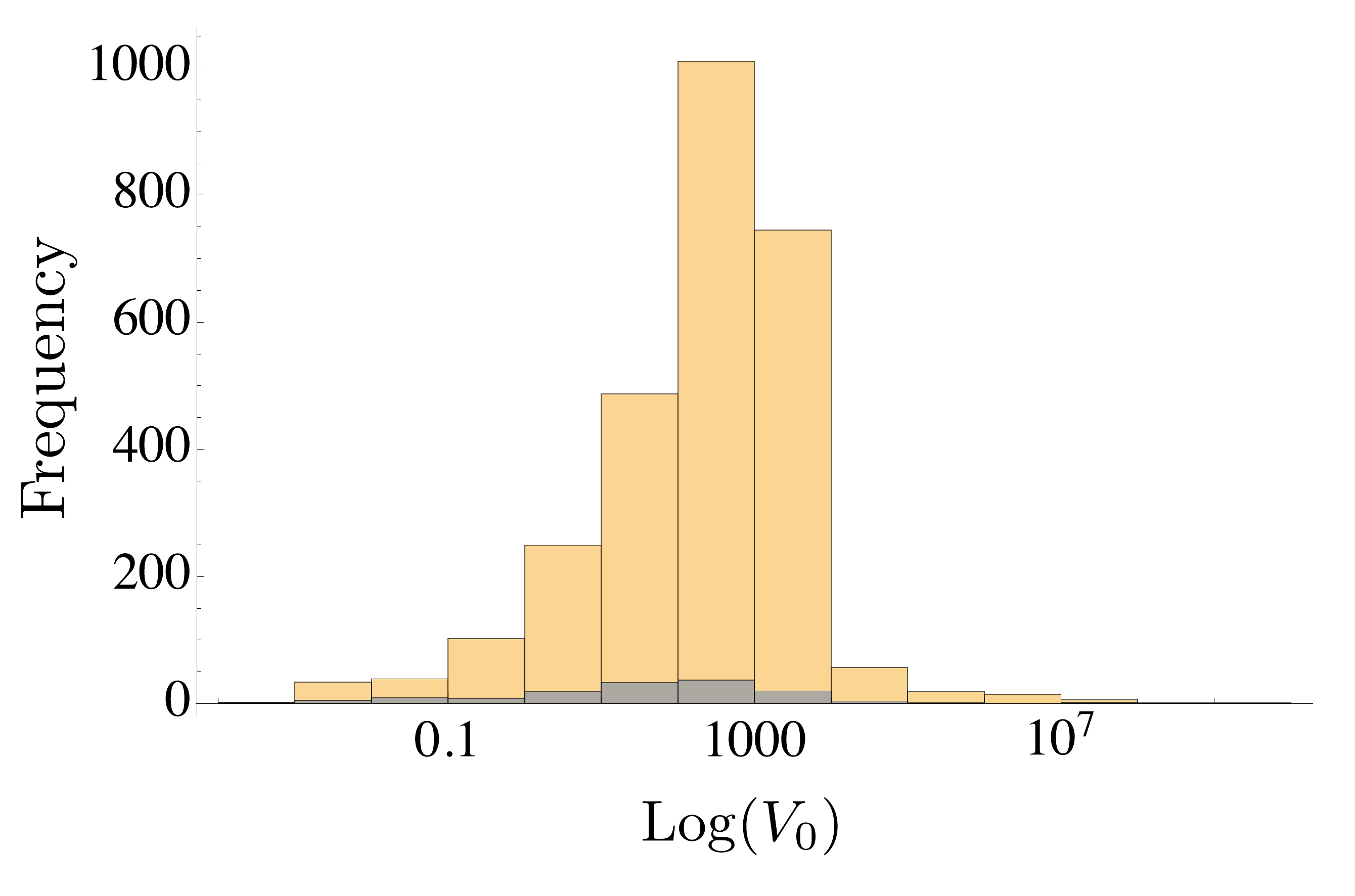}\Lb  \\ 
\includegraphics[scale=0.28]{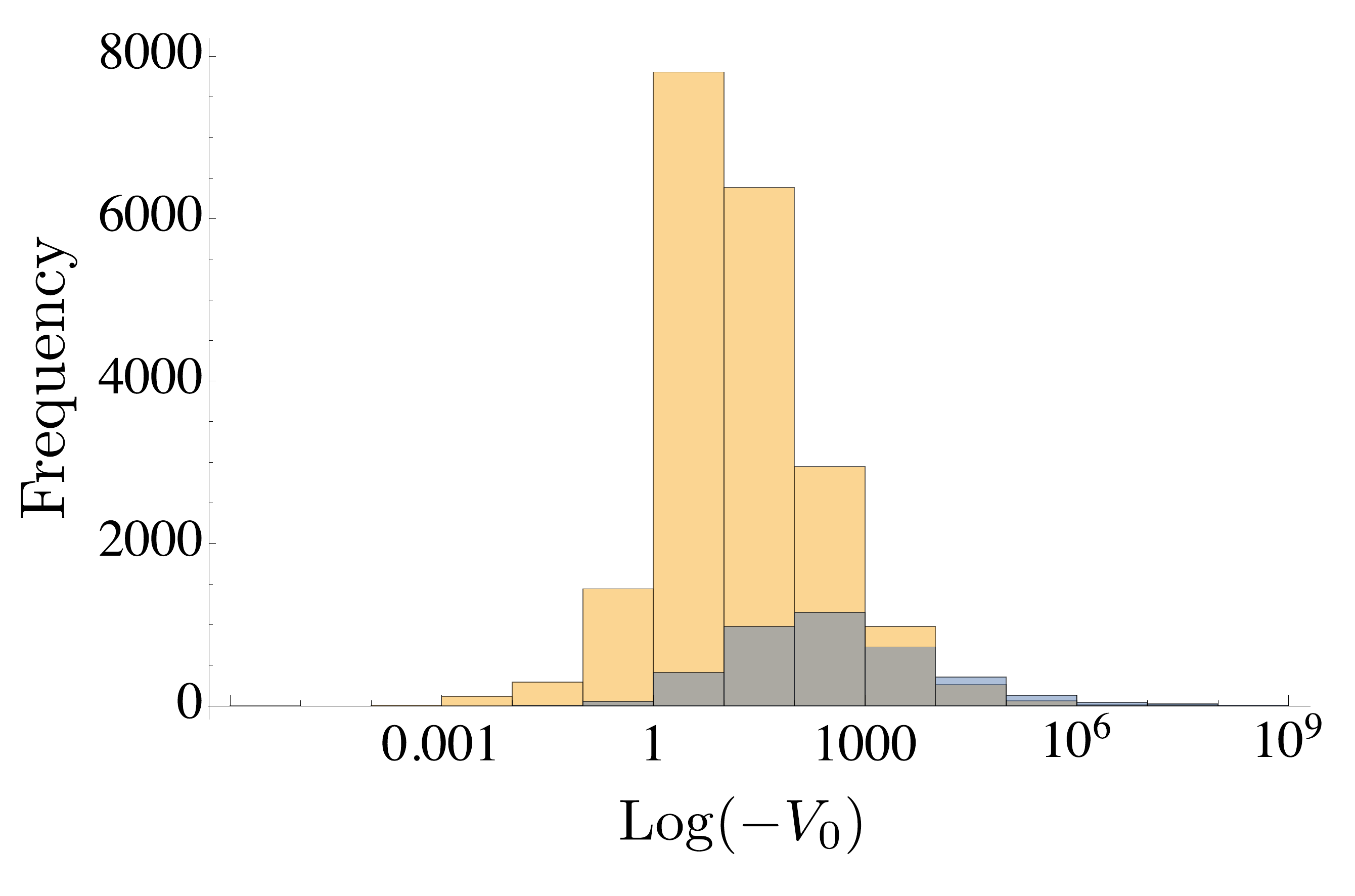}\Lc  \includegraphics[scale=0.28]{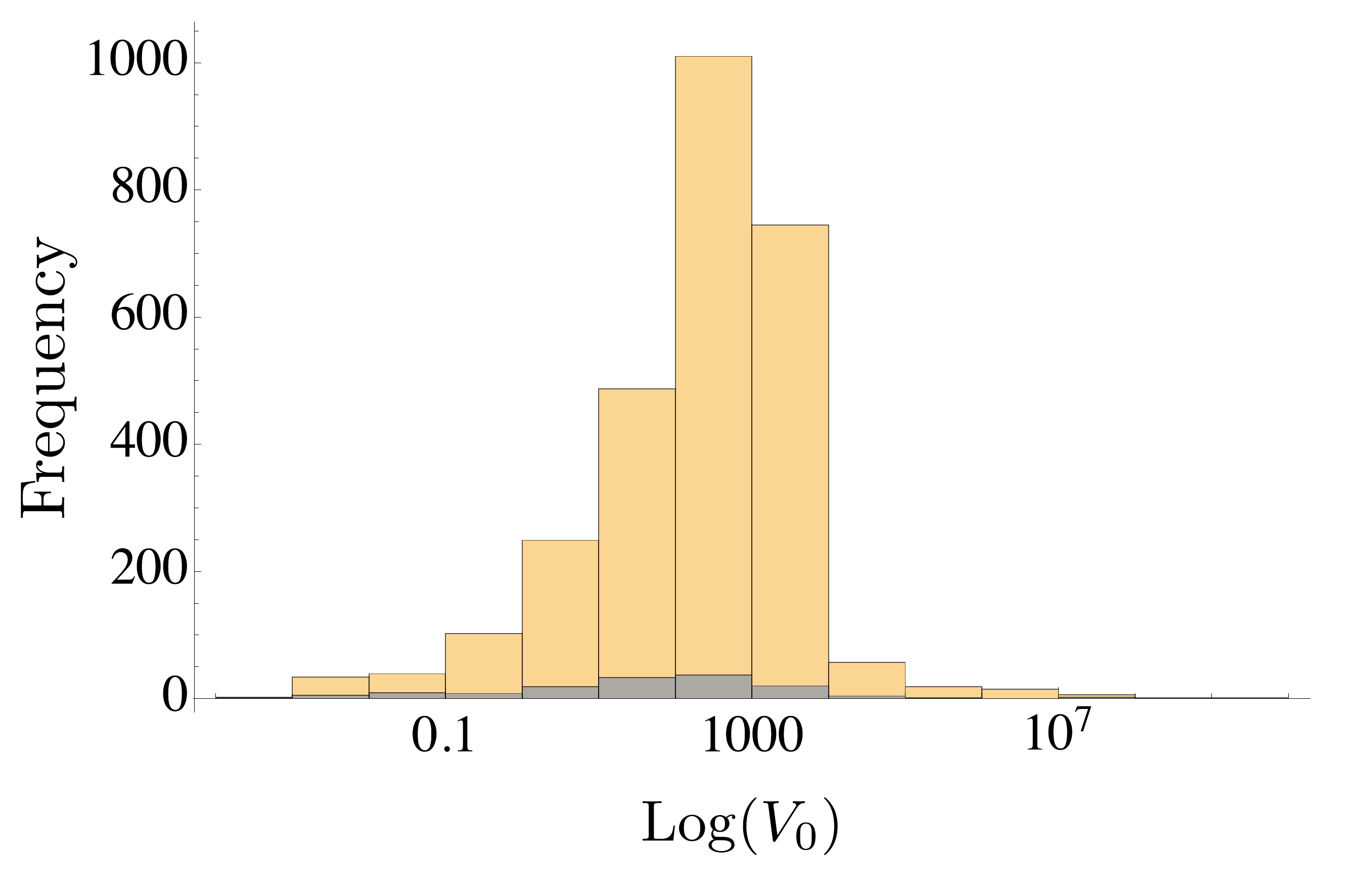}\Ld \\ 
   \caption{\emph{Histogram of critical points with random fluxes obtained by the ANN+GA (yellow bars) and those randomly found by the GA (blue bars). Intersection appear as grey bars. (a) AdS vacua, Criterion II, (b) dS vacua, Criterion II, (c) AdS vacua, Criterion I, (d) dS vacua, Criterion I. Notice that the figure a) differs from the others in that
   the amount of solutions found by the GA alone and the ANN+GA are comparable, this reflects the fact that in this case the training set for the ANN was smaller. In contrast with Figure \ref{fig:varvac}, these histograms include the cases where at least exists one tachyonic state.}}
   \label{fig:varhier1}
\end{figure}

The ANN flux classification improves our capacity to find vacua and in consequence to explore the String Landscape or the Swampland. This follows from the analysis plotted in Figure \ref{fig:varhier1} where we show the number of vacua, stable or not, versus the value of the scalar potential at the critical point. We notice that for the case of AdS, the number of vacua is increased by the use of the ANN compared to those obtained by GA for the Case I. However the same is not true for Case II. On the other hand, the number of dS vacua  increases by the use of the ANN in both cases, although neither of them contain a stable dS vacuum. See Figure \ref{fig:varhier1} for more details. By looking at the order of magnitude on the number of vacua found by the use of the ANN, we conclude that Case I is much more efficient than Case II. \\

\subsection{Case B. Hierarchy on Fluxes}

\subsection*{ANN Training}

In this case the ANN is trained by an input of flux configurations with a clear hierarchy on their integer values. This hierarchy means that the integer values parameterizing one of the sectors, e.g. NS-NS, R-R or nG 
are between one and four orders of magnitude bigger than the fluxes in the other sectors. 
As in Case A, all flux configurations satisfy the usual constraints of tadpole cancellation  and Bianchi identities with no D-branes.
We explore 3 hierarchies among the fluxes: $f,h \gg b$, $h,b \gg f$ and $f,b\gg h$. The inequalities
imply that all the flux components of one kind differ by at least one order of magnitud from all the flux
components of the other kind (i.e. for the first type $\forall_{i,j,k} f_i \gg b_k,\, \,  h_j \gg b_k$).  \\

A hierarchy on the integer values associated to all fluxes in turn establishes a hierarchy on the masses associated to the moduli.
This is, if we take for example that R-R fluxes to be larger than the others $f\gg h, b$, we expect in this model, that the complex structure modulus would be the heaviest modulus $M_U \gg M_S,M_T$ \cite{CaboBizet:2019sku}. 
Next we write the expected hierarchies between the moduli masses that are obtained by setting one of the explored
hierarchies among the fluxes:
\begin{eqnarray}
{\rm Case}\,\,{\rm K:} & f,h \gg b  \, \,  \, \,  &\rightarrow  \, \,  \, \,  M_U,M_S \gg M_T,\nonumber\\
{\rm Case}\,\,{\rm CS:}  &h, b \gg f\, \,  \, \,   & \rightarrow \, \,  \, \,  M_S,M_T \gg M_U, \\
{\rm Case}\,\,{\rm AD:} &  f , b\gg h\, \,  \, \,   & \rightarrow \, \,  \, \,  M_T,M_U \gg M_S. \nonumber
\end{eqnarray}

The classification as in the previous case is done by demanding the ANN to identify flux configurations which generate a scalar potential with a stable critical point\footnote{We have discarded an analisys for Criterion II once we conclude it is not efficient when applied to hierarchical fluxes.}. Since the flux configuration presents a hierarchy,  all the critical points are also related to a spectrum with a lightest moduli.
Notice that for this case \textit{we are not training the ANN to find critical points with a positive value for the scalar potential}. This follows from our experience in case A in which the dS criterion (Criterion II) did not produce much more vacua as  desired. In Figure \ref{fig:varhier} and Figure \ref{fig:varcphier} stable and critical points are analyzed.\\


\subsection*{Results}

\begin{figure}[!htbp]
   \centering
\includegraphics[scale=0.28]{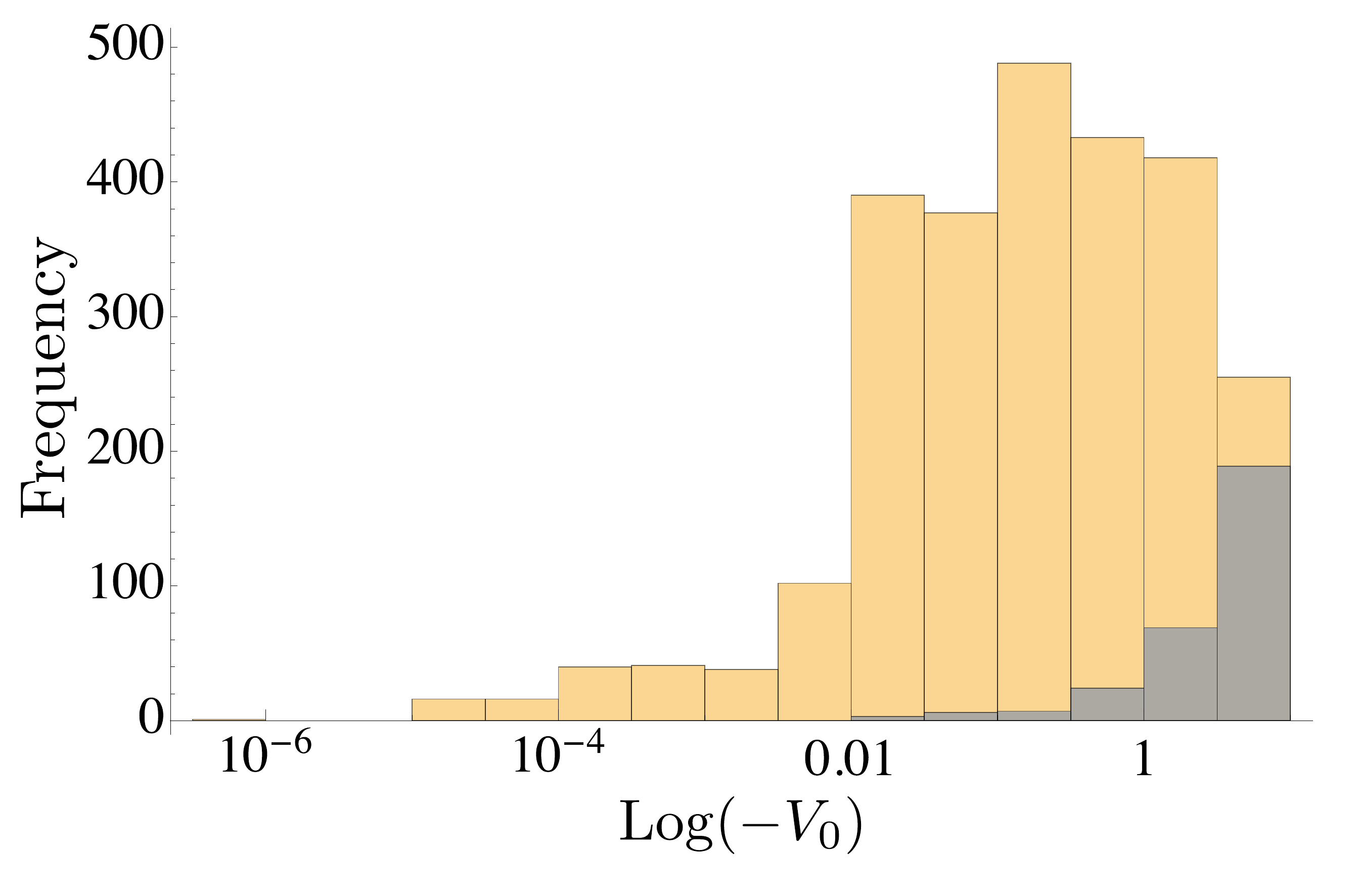}\Laa \includegraphics[scale=0.28]{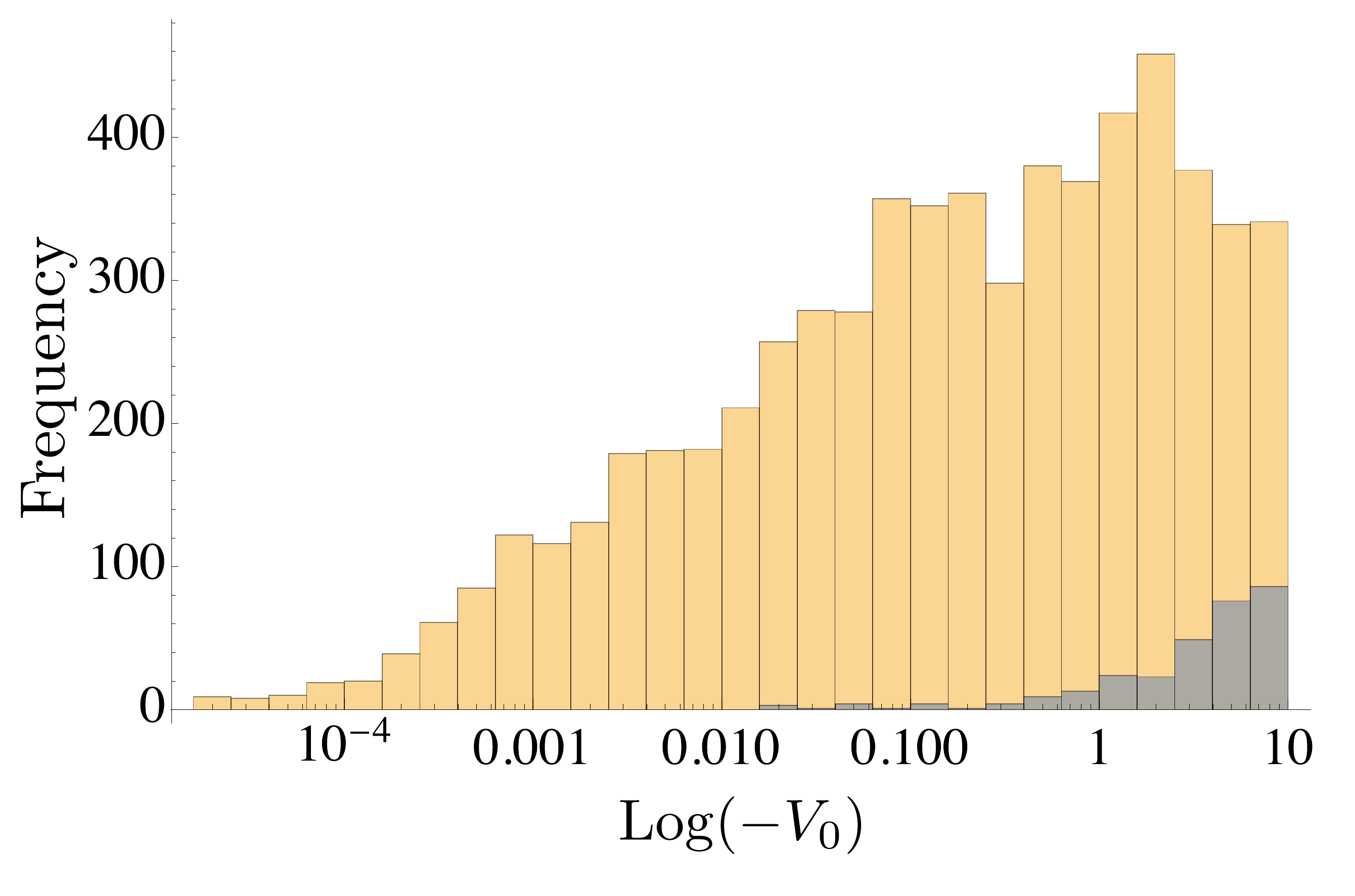}\Lbb
   \\ \includegraphics[scale=0.28]{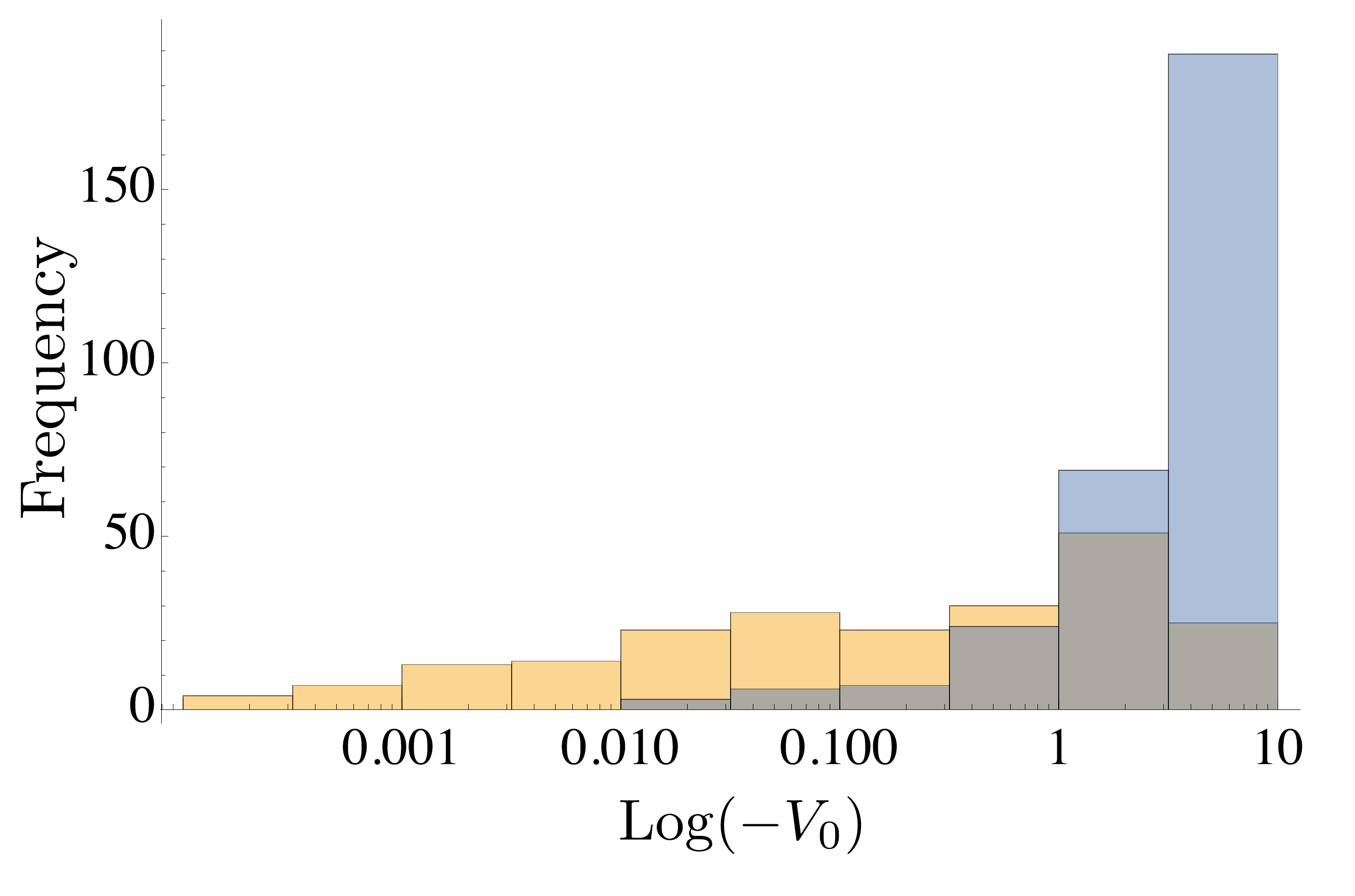}\Lcc
   \caption{\emph{Distributions of vacua obtained for hierarchical fluxes:  (a) K\"ahler (K), (b) complex structure (CS) and (c) axio-dilaton( AD) as the lightest modulus. 
   The critical points obtained by the ANN+GA are given in yellow bars, and those randomly found by the GA are given in blue bars. This data represents the free tachyon spectrum classification of fluxes with hierarhcy.}}
   \label{fig:varhier}
\end{figure}

The histograms obtained after ANN's classification are shown in Figure \ref{fig:varhier}. As observed,  selecting a specific hierarchy on the flux configuration affects the distribution of vacua: 
\begin{itemize}
\item If we take, for instance, both R-R and NS-NS larger than nG fluxes ($f,h \gg b$), we obtain the lightest mass to be that of the K\"ahler modulus $M_T$. 
In this case we notice a clustering of 
the number of stable vacua around a given value for the cosmological constant well below the peak obtained for randomly selected fluxes,
with a mean value of the cosmological constant lower than its value on the randomly selected vacua. 

\item If we take the Complex Structure 
as the lightest modulus ( $h , b\gg f$), we observe an increase in 
the number of stable AdS vacua  with a greater dispersion. However, for the case in which the lightest modulus is the Axio-Dilaton ( $f , b\gg h$), we do not notice an improvement on the amount of stable vacua in relation with a random flux configuration input.
\end{itemize}

\begin{figure}[!htbp]
   \centering
 \includegraphics[scale=0.28]{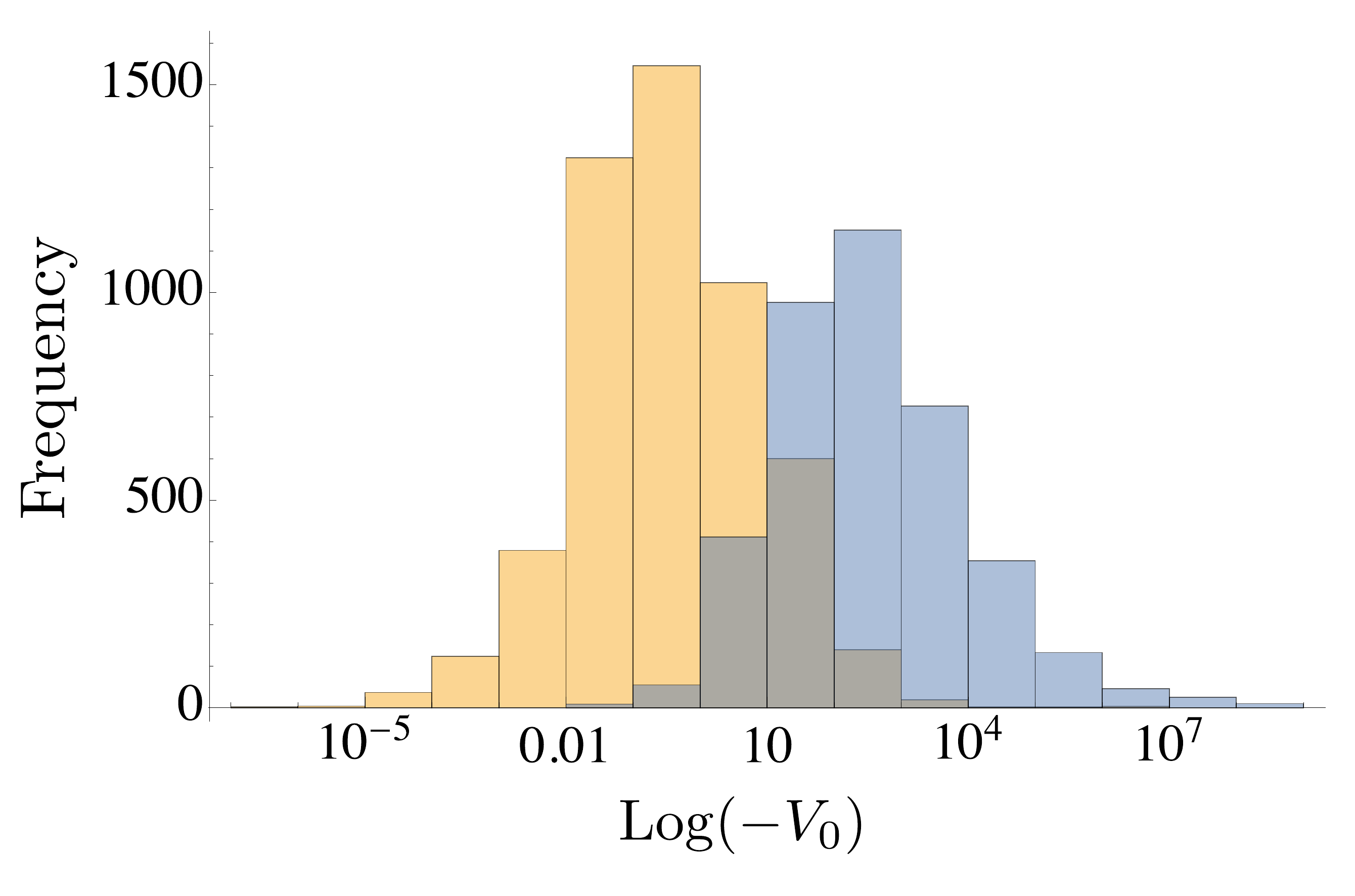}\La \includegraphics[scale=0.28]{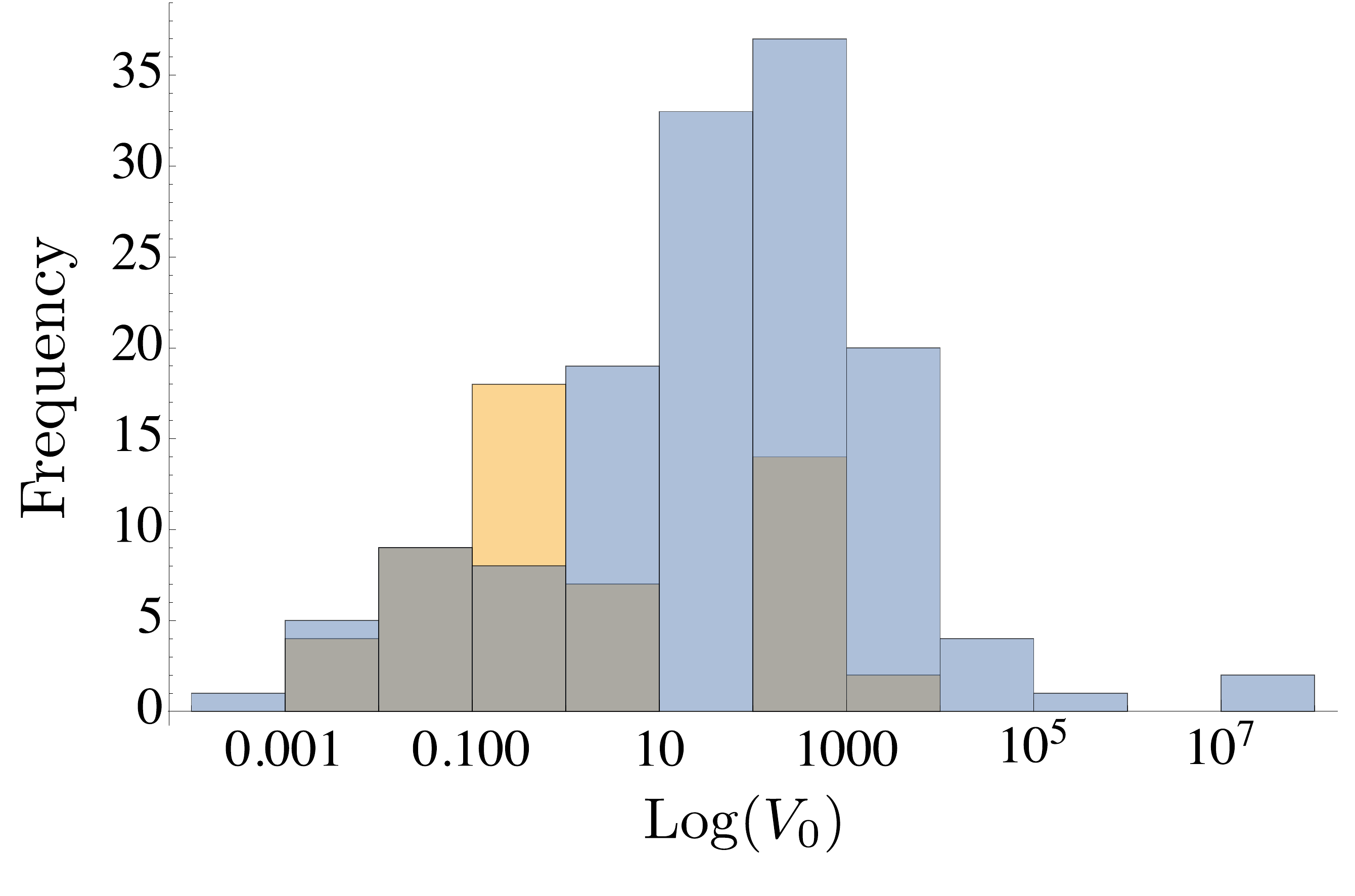}\Lb \\ 
\includegraphics[scale=0.28]{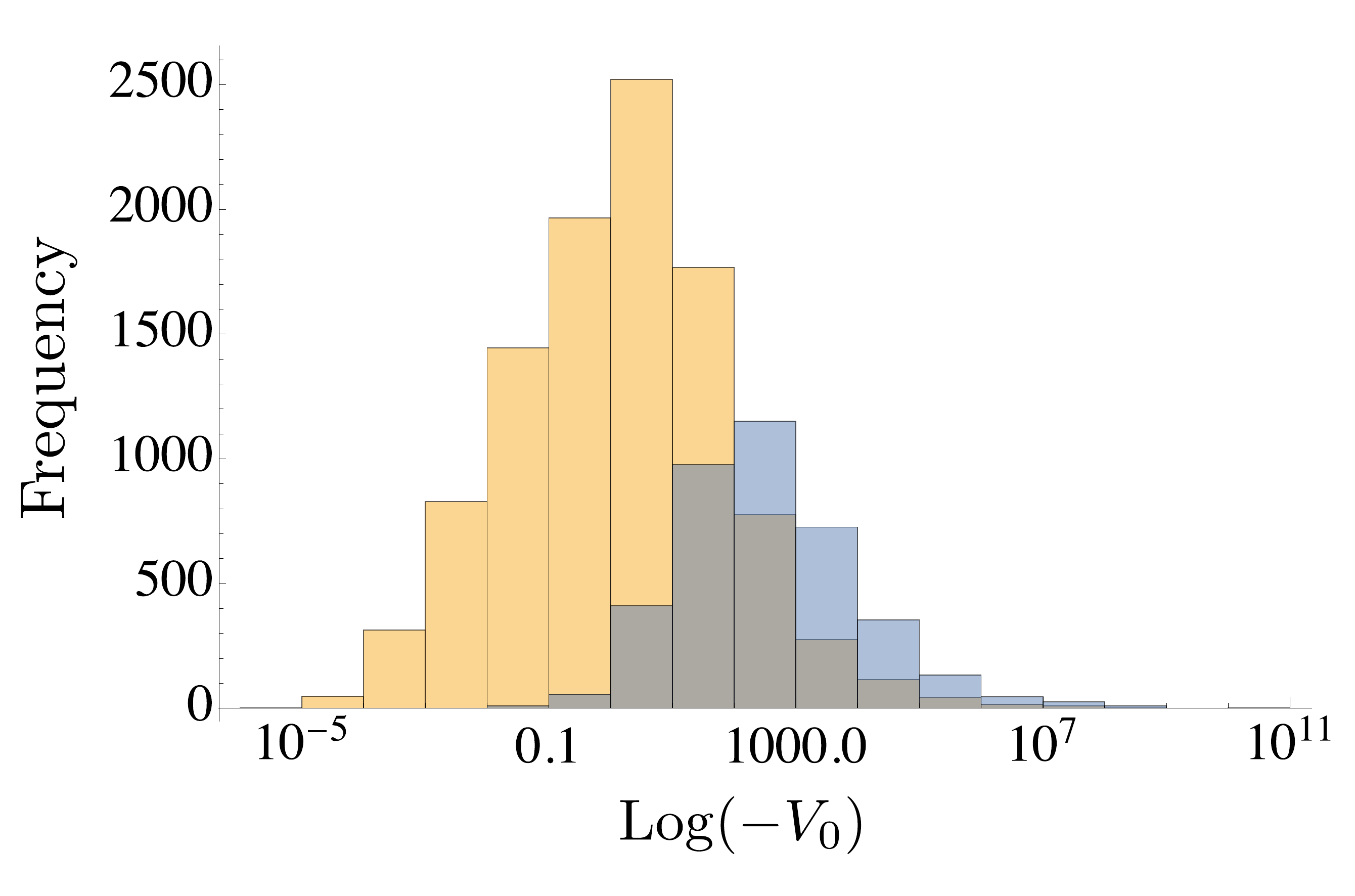}\Lc \includegraphics[scale=0.28]{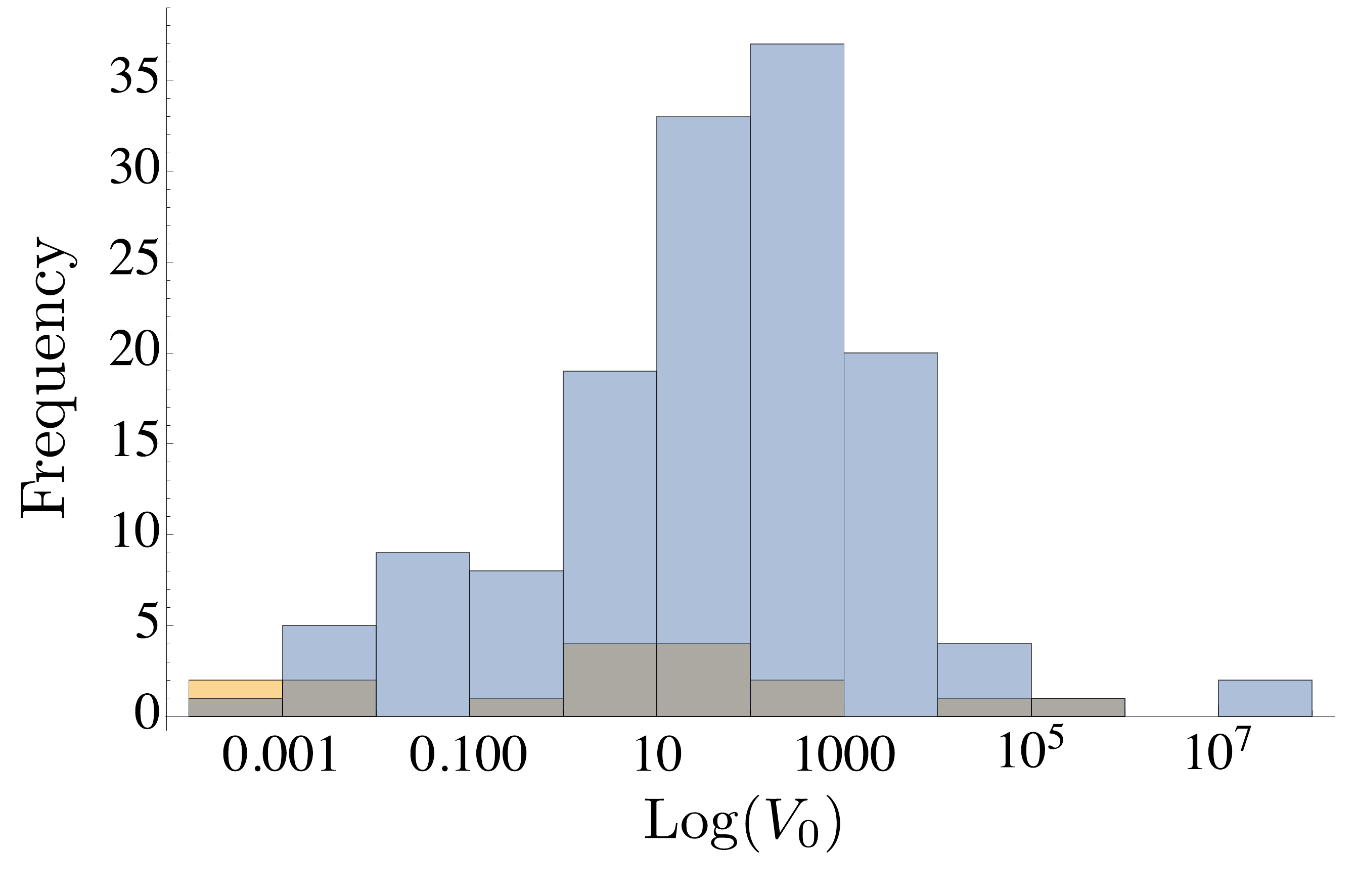}\Ld \\
 \includegraphics[scale=0.28]{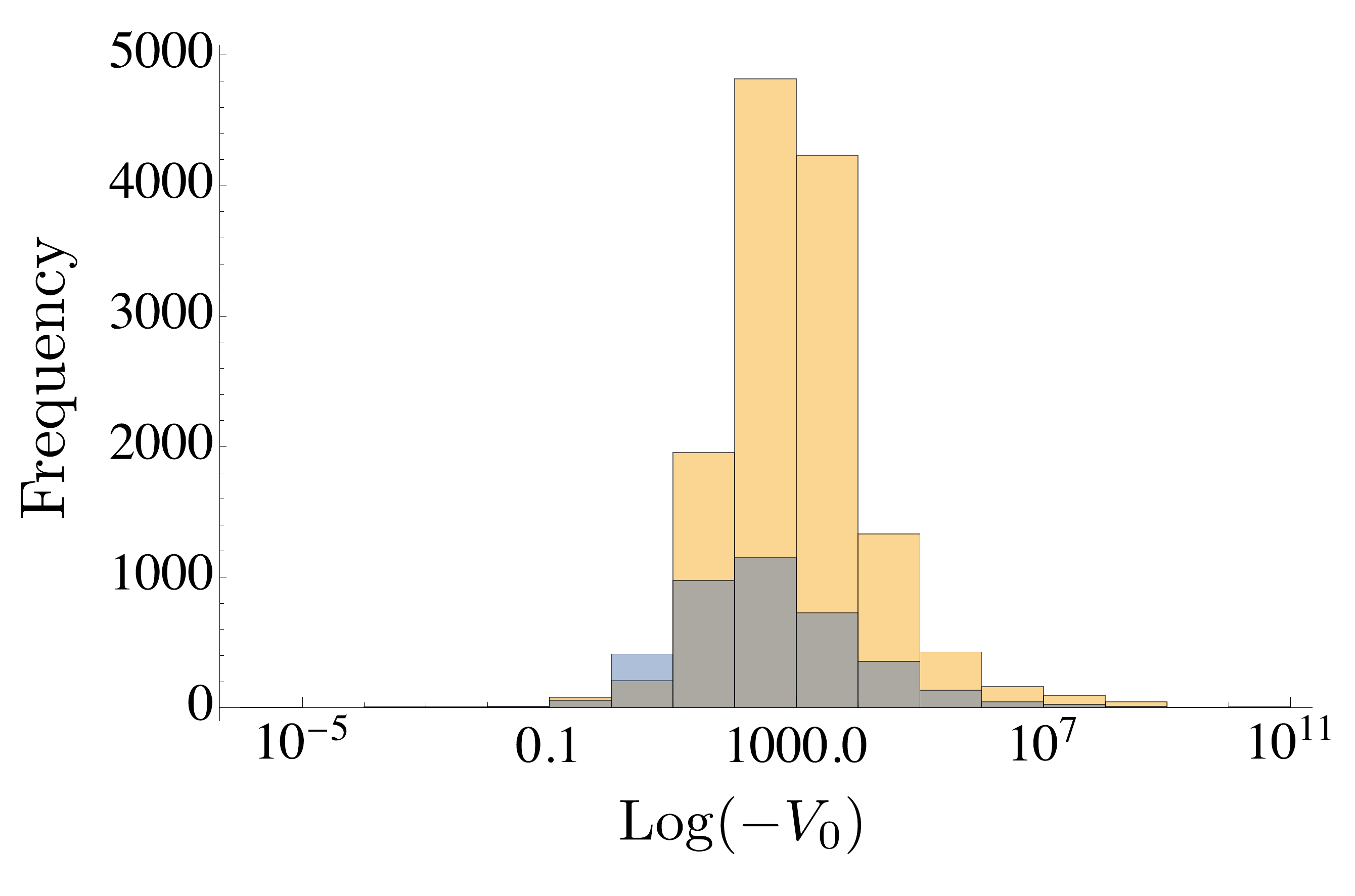}\Le \includegraphics[scale=0.28]{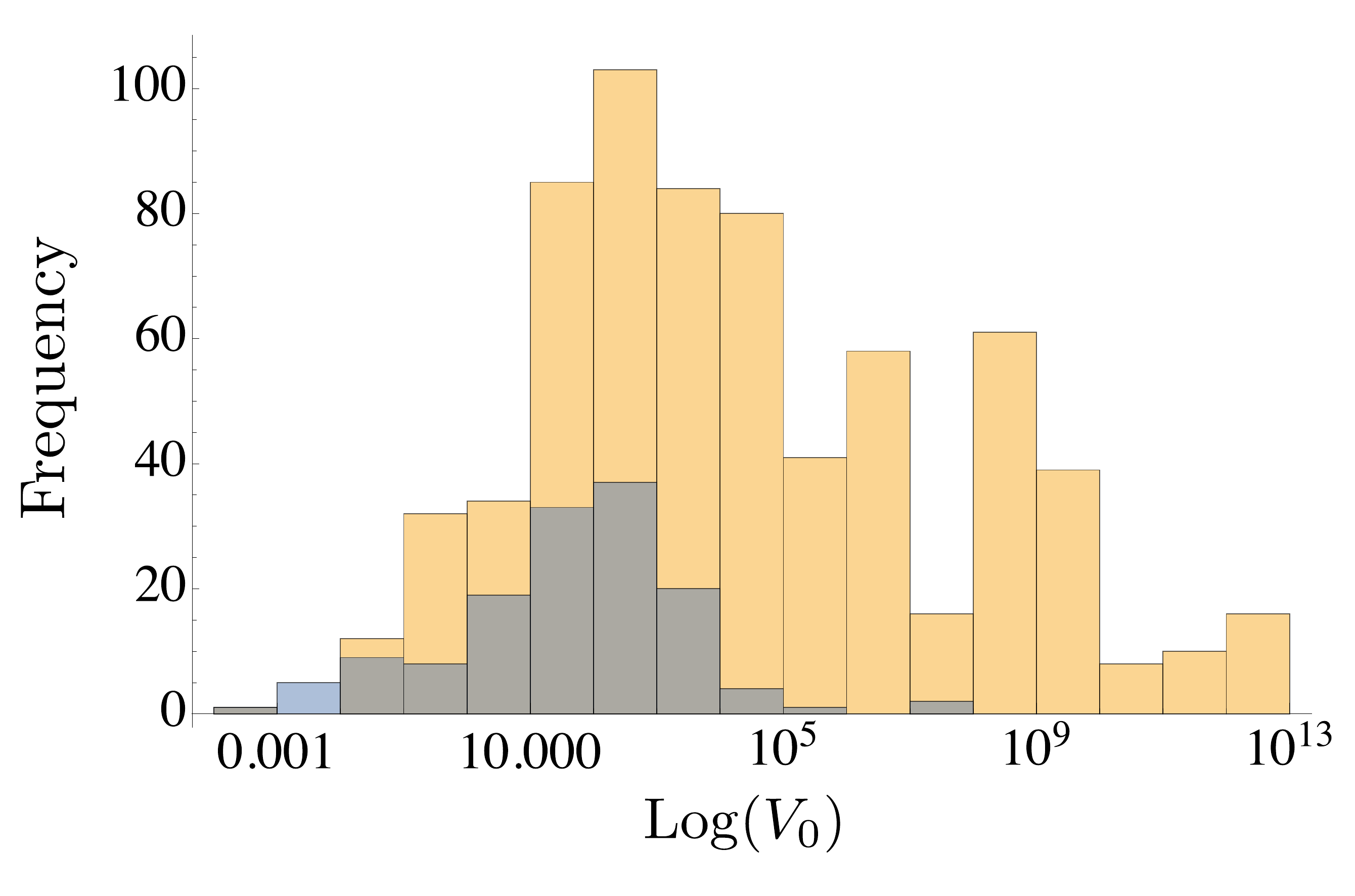}\Lf
   \caption{\emph{Distributions of Vacua for hierarchical fluxes. (a) AdS vacua, case K, (b) dS vacua, case K, (c) AdS vacua, case AD, (d) dS vacua, case AD, (e) AdS vacua case CS, (f) dS vacua, case CS. All dS vacua are unstable.
   The critical points obtained by the ANN+GA are given in yellow bars, and those randomly found by the GA are given in blue bars. In contrast with Figure \ref{fig:varhier} these histogram includes the vacua that contains at least one tachyon.}}
   \label{fig:varcphier}
\end{figure}

In Figure \ref{fig:varcphier} we present the corresponding  histograms related to different hierarchies on the moduli masses. 
Notice that for all cases the histograms seem to follow a normal distribution. Figures (a) and (b) indicate the distribution of vacua for the case in which the K\"ahler modulus is the lightest one $M_S,M_U\gg M_T$ (case K) against the value $-$negative or positive$-$  of the scalar potential at that point. Figures (c), and (d) correspond to the case in which the axio-dilaton modulus is the lightest one $M_T,M_U\gg M_S$ (AD case) ; whereas Figures (e) and (f) refer to
the case in which the complex-structure moduli is the lightest one $M_T,M_S\gg M_U$ (CS case). \\

The ANN classification shows a greater abundance of AdS critical points for this Case B than for Case A. Besides, the critical points for the K and CS  cases respectively, 
have a mean value for the scalar potential lower than the value on the AD case. Conversely, the abundance of dS critical points is reduced in the K and CS cases 
in comparison with Case A. \\

\section{Surveying the Landscape of Vacua}
\label{sec:tres}

Upon correlation of different features for the vacua we obtained, we draw three important observations, which we present in order. 

\subsection{Perturbative regime is associated to a small minima of the scalar potential}
 \label{sec:AD}

\begin{figure}[htbp]
   \centering
    \includegraphics[scale=0.60]{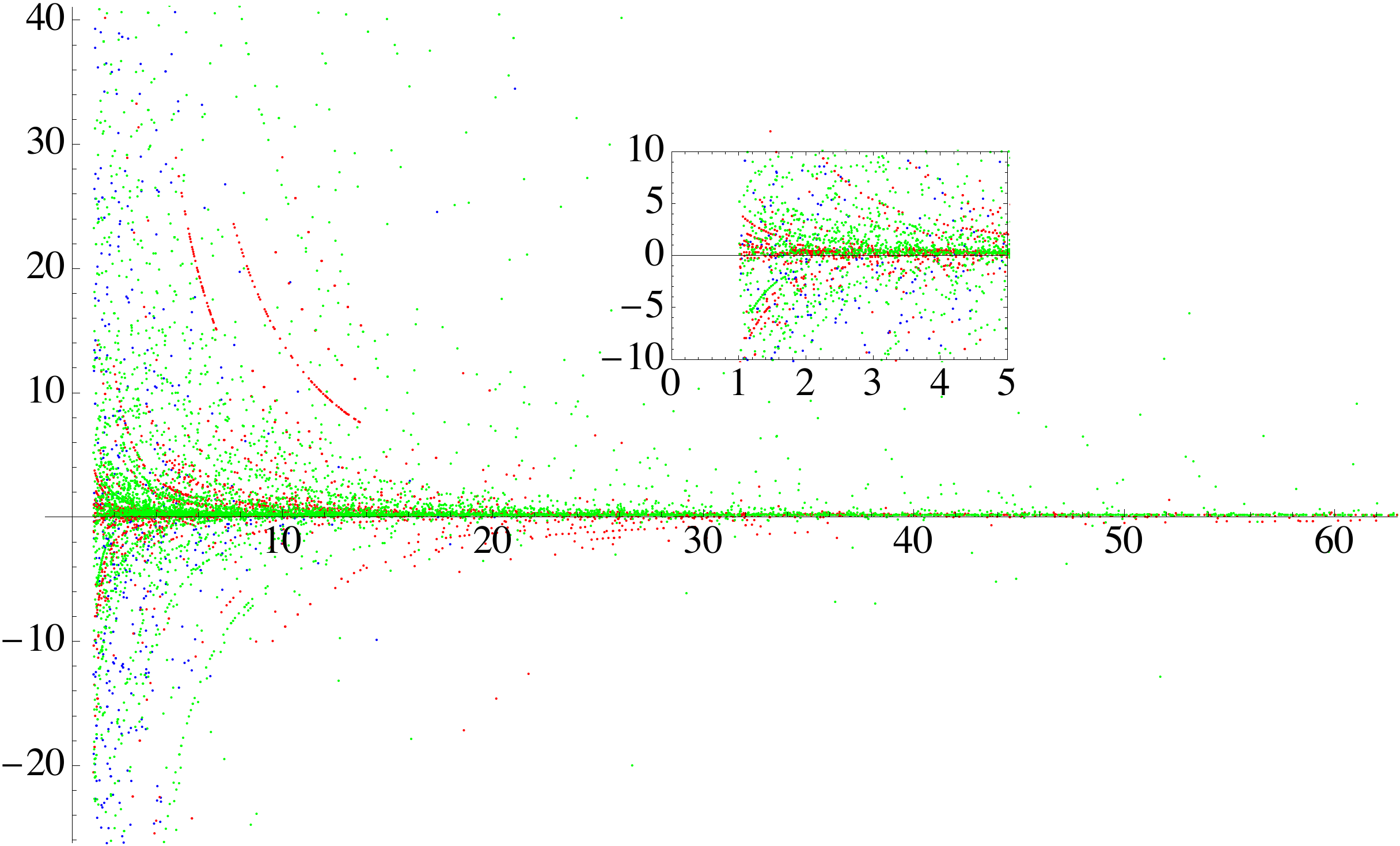} 
    \begin{picture}(0,0) 
   \put(-10,102){$\text{Re}\,S$}
   \put(-400,250){$\Lambda$}
  \end{picture}
   \caption{\emph{Value of the scalar potential versus the string coupling at the critical point for all analyzed cases produced by the ANN+GA. Red and Blue points correspond to vacua classified in Case A, while yellow and green dots are related to Case B where a hierarchy of the flux configurations is assumed. It is observed that the smallest the string coupling, then the smallest value for the cosmological constant $\Lambda$.}}
   \label{fig:cp}
\end{figure}
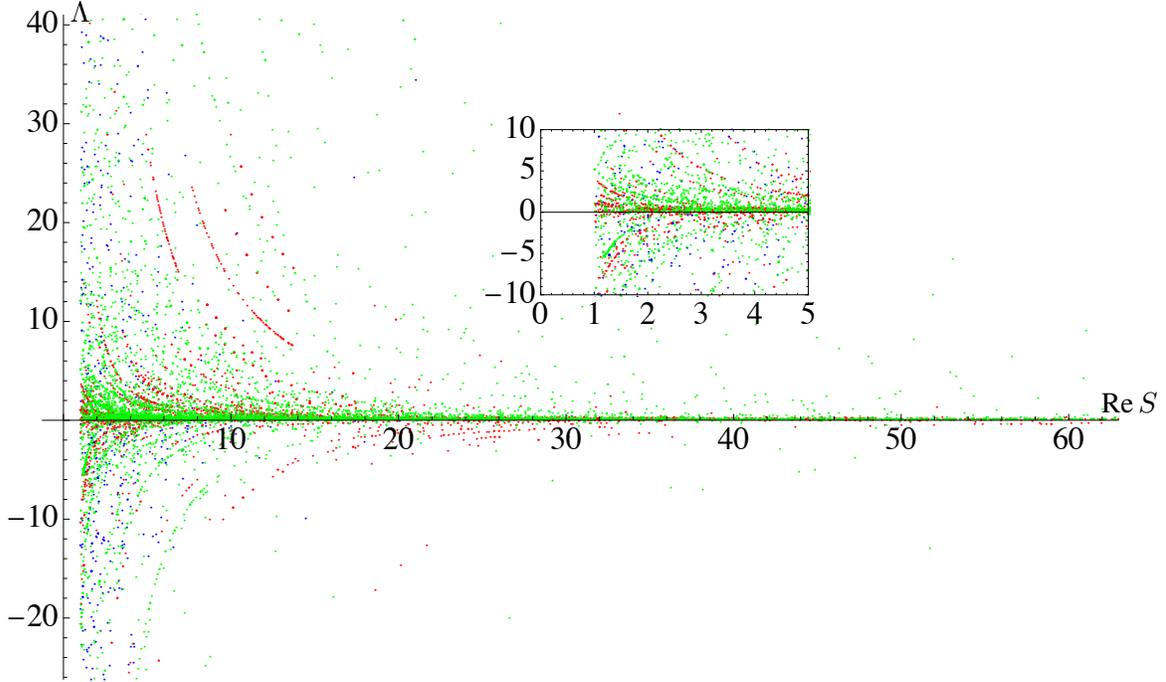

A careful comparison of critical points shows that the largest values of the scalar potential at the corresponding critical point are related to  non-perturbative regime ($Re\, S \ll 1$), and thus cannot be trusted. This can be seen in Figure \ref{fig:cp} where we have plotted all AdS and dS vacua (not necessarily stable) obtained by the ANN against the string coupling value (real part of the axio-dilaton at the critical point). \textit{We therefore observe that those flux configurations associated with very small values for the string coupling, i.e., describing an effective perturbative model, are related to small values for the cosmological constant, suggesting a relation of the form $\Lambda = \pm \exp \left( - \text{Re}\, S \right)$}.
%
%

\subsection{Compatibility with the Refined dS conjecture}

\begin{figure}[htbp]
   \centering
    \includegraphics[scale=0.60]{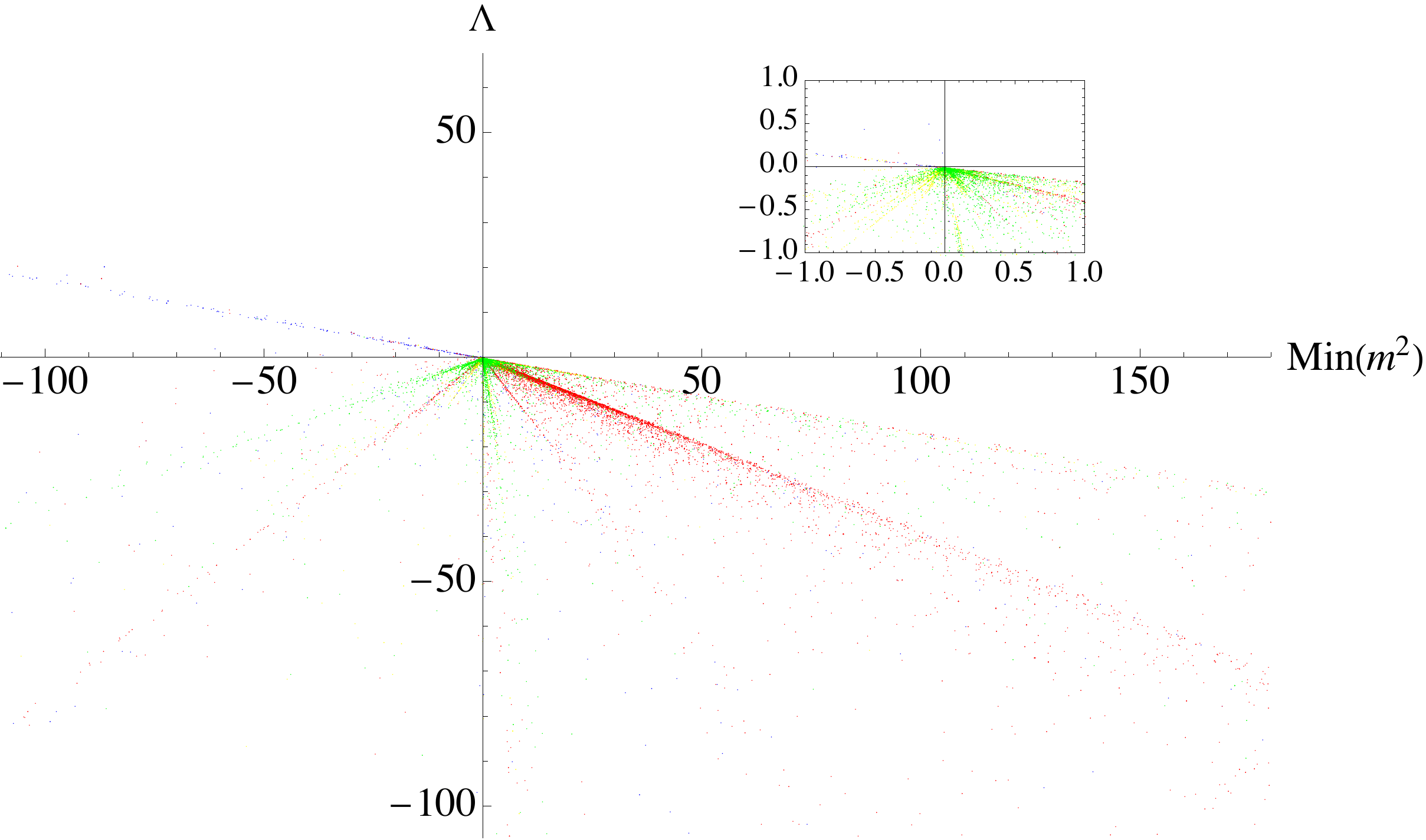}  
   \caption{\emph{Distribution diagram of the values of the extrema for the scalar potental RdSC. Red and blue points represent critical points obtained through the classification of randomly selected flux configurations (Case A) whereas green and yellow points correspond to critical points obtained by assuming hierarchical fluxes (Case B).}}
   \label{fig:refdS1}
\end{figure}

The smallest eigenvalue of the $\nabla_i \nabla_j$ operator, denoted $\text{min} \,\nabla_i \nabla_j V$, corresponds to the mass of the lightest modulus (which in the case of an unstable vacuum is tachyonic). 
Using the vacua distribution of the values of the potential at the critical point ($\Lambda$) versus the smallest modulus mass, we graphically observe that the vacua obtained populates only a half of that plane: essentially all the data lies below the line $V= -\frac{1}{c'} \text{min} \, m^2+ c''$ for some for $c'' < 0$. As mentioned before, 
the slope of the line is related to $c'$ parameter. In Figure \ref{fig:refdS1}, vacua obtanied in Case A are represented by red  (Criterion I) and  blue (Criterion II) points, while green (case K) and yellow (case CS) points represent those obtained 
in Case B.\\
%

From this analysis we conclude the following:
\begin{itemize}
\item The dispersion shows a structure in the vacua corresponding to straight lines. Different vacuum solutions in the same line belong to a set of fluxes related to a particular solution of Bianchi Identities and Tadpole conditions.
\item Straight lines do not pass through the origin, instead they are displaced a small amount parametrized by $c''$. This is related to the fact that we look for solutions in which the second derivative for the scalar potential is different from zero.
\item The hierarchies move the critical points towards the origin. This implies that, by demanding a hierarchy on the flux configuration input, the minima of the scalar potential becomes smaller, and according to our previous observation \ref{sec:AD} a smaller string coupling is also obtained.
\item Notice that this classification indeed reproduces the expected plot shown in Figure \ref{fig:RdSC}, indicating not only the absence of stable dS vacua, but also the absence of some stable AdS and the presence of some unstable dS limited by a straight line. 
\item
The vacuum points lie very close to the origin in Figure \ref{fig:refdS1} representing critical points with a small negative vacuum energy and with a small value for $m_{ij}^2$, indicating that very close to the minimum there could be conditions on the scalar potential for which the AdS scale conjecture could be violated. It is then important to study how probable is to to find such solutions.

\end{itemize}

\subsection{AdS Scale Separation}
\begin{figure}[!htbp]
   \centering
  \includegraphics[scale=0.28]{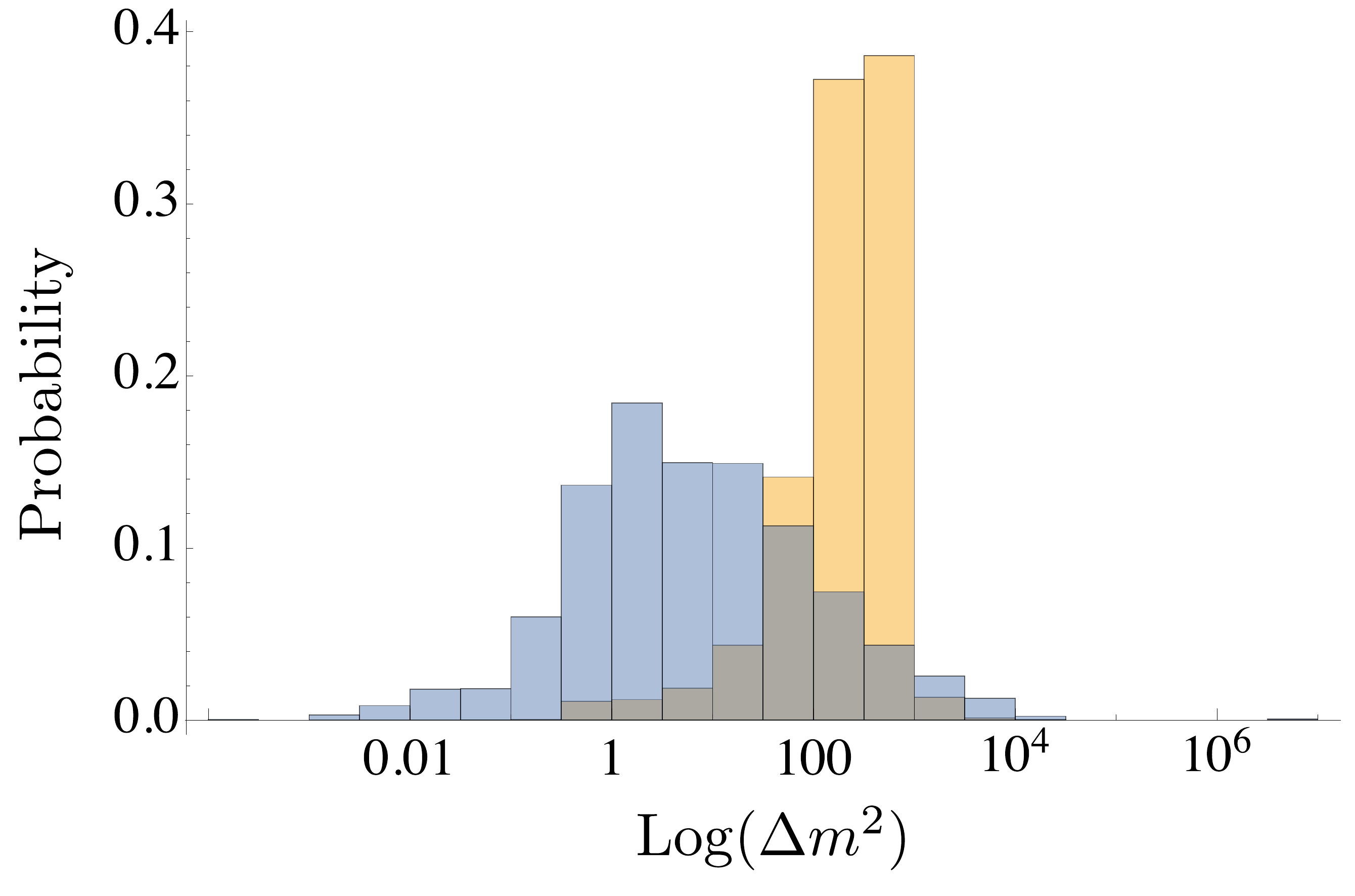}\Laa \includegraphics[scale=0.28]{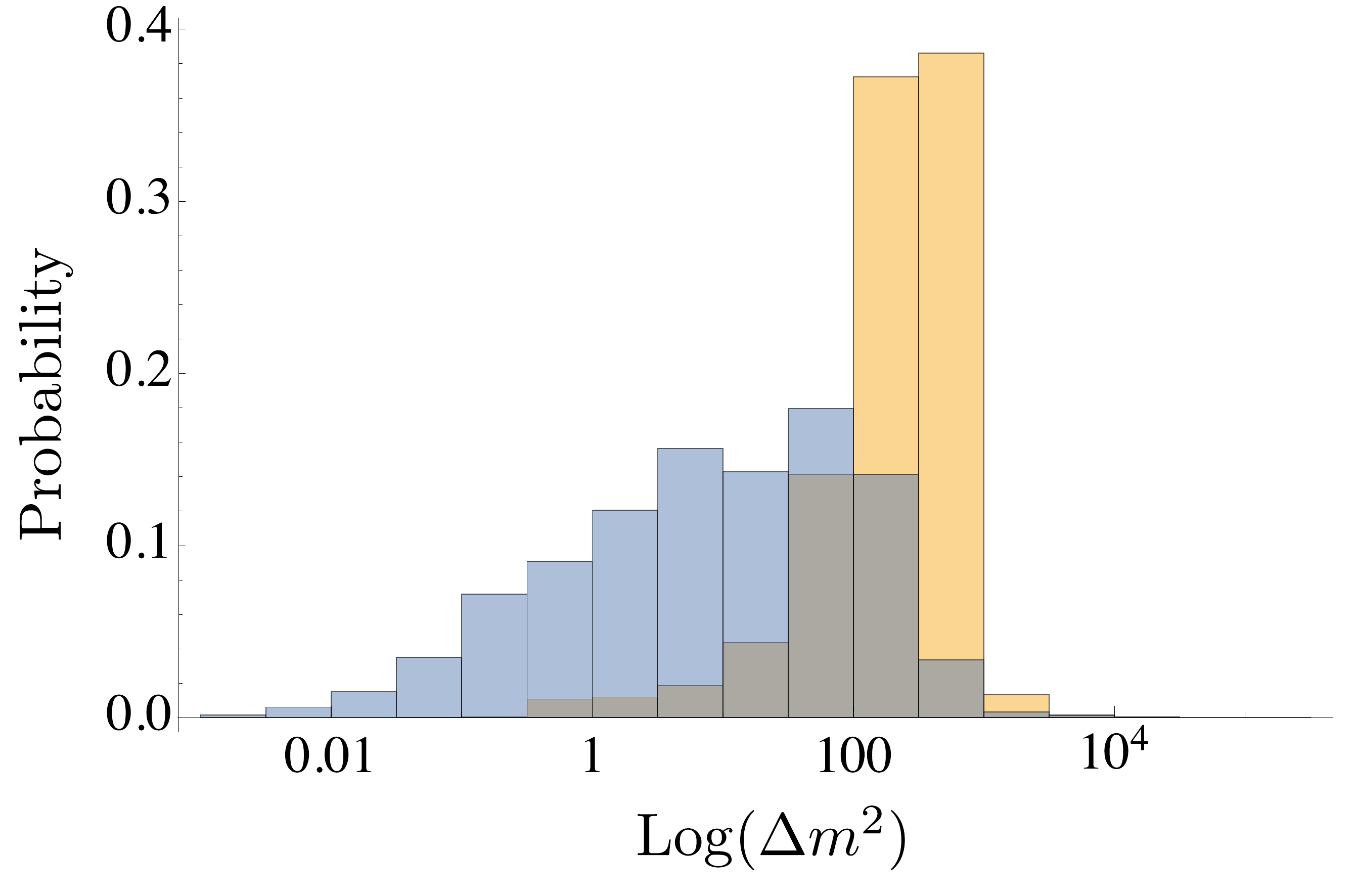}\Lbb \\ 
 \includegraphics[scale=0.28]{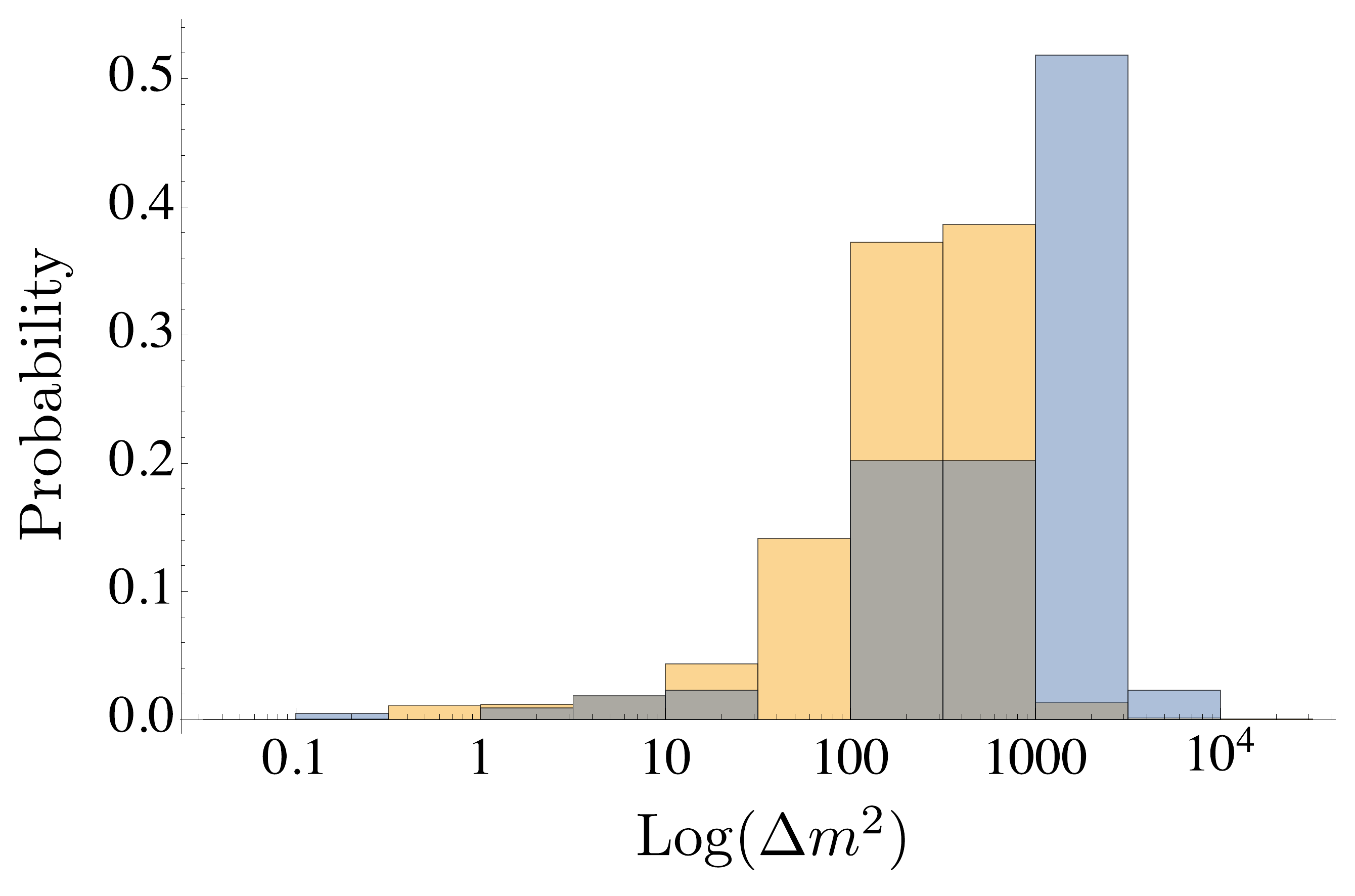}\Lcc 
      \caption{\emph{Probability histograms for the scale separation between the heaviest and the lightest mode. Different histograms correspond to  the case when a given modulus
is the lightest: a) K\"ahler (T), b) Complex structure (U)  and c) Axio-dilaton (S). Yellow bars refer to hierarchical flux configurations while blue bars refer to non-hierarchical ones.}}
   \label{fig:hierarTSU}
\end{figure}

Let us now classify  the scale separation between stable AdS vacua  $\Lambda_{\text{AdS}}$ 
and the squared mass corresponding to the lightest modulus for all models constructed from a Case A configuration. 
This study allows us to directly see, as shown in Figure \ref{fig:hierarTSU}, that by using a configuration of hierarchical fluxes it is more probable to find a hierarchy among moduli masses. Limited to our model we can say that the most probable scenario involves a maximum difference of masses of order of magnitude 3 
where the difference is given by
\eq{
\Delta m^2 = \text{max}\, m^2 - \text{min}\, m^2 \,.
}
Notice that an exponential $\Delta m^2$ as present in a KKLT model is discarded in our case, probably due to the fact that we are considering a hierarchy among fluxes of an order of magnitude between 1 and 4 
which in turn is a consequence of Bianchi and Tadpole constraints \cite{Betzler:2019kon}.\\

The AdS swampland scale conjecture asserts that it is not possible to separate the size of the AdS space and the mass of its lightest mode beyond a certain limit, this is
\eq{
\left( \text{min} \, m^2 \right) L^2_{\text{AdS}} \leq c \,,
}
where $c$ is constant of order 1, and  $L^2_{\text{AdS}} \sim \Lambda^{-1}_{\text{AdS}}$. This conjecture is motivated from the point of view of the KKLT scenario, in the sense that any uplifting mechanism (from a supersymmetric stable vacua) does not destabilize the K\"ahler moduli as far as the potential well is parametrically narrow in comparison with the 
energy gap that needs to be filled by the uplifting mechanism. For the KKLT scenario, indeed this criteria is not fulfilled and thus it raises the question of its validity \cite{Gautason:2018gln}. \\

\begin{figure}[!htbp]
   \centering
   \includegraphics[scale=0.29]{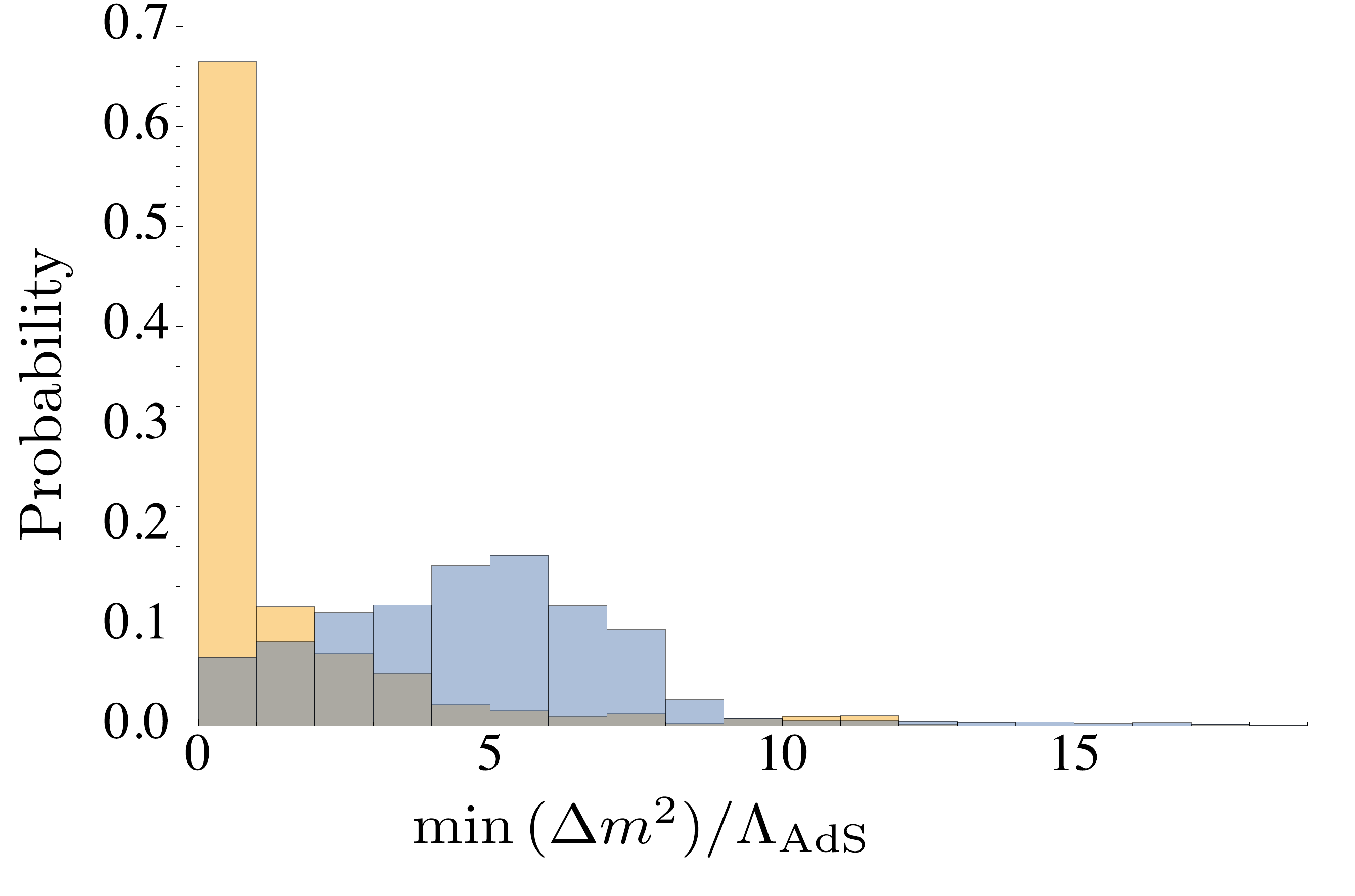}
      \caption{\emph{Histogram of the scale separation between the lightest modulus and the corresponding value of the cosmological constant in Planck units ${\rm min}(m^2/\Lambda_{\rm AdS})$. Yellow bars correspond 
      to non-hierarchical fluxes while blue bars correspond to fluxes with a given hierarchy.}}
   \label{fig:AMSSC}
\end{figure}

We analyze this conjecture for our simple model (see Figure \ref{fig:AMSSC}) and we observe that the effect of  using a hierarchical flux configuration solution implies an increase on  
the probability to find a scenario in which the scale between the lightest modulus and the size of AdS space be in the lower bound. 
Thus as argued by \cite{Gautason:2018gln}, any attempt to uplift the AdS vacua may destabilize the lightest modulus. However, for fluxes without hierarchy (yellow bars) it is observed that the mean value of the vacua is around 1 saturating the bound. Therefore we conclude that a hierarchical flux configuration leads us to scenarios in which the ratio ${\rm min}\, m^2/\Lambda_{\rm AdS}<1$ 
, which according to the conjecture of AdS scales, would produce an instability if uplifting to dS. \\

In summary, \textit{by assuming a hierarchy on the flux configuration among different sectors (NS-NS, R-R and NG) it is more probable for the generated vacua to have small values for the vacuum energy, a small value for the string coupling. Also, scenarios constructed with non-hierarchical fluxes exhibit a higher probability for the lightest modulus to be much larger than the cosmological constant, which according to a recent conjecture, makes impossible to uplift to a dS vacuum.}\\

However, among all possible flux configurations, having a hierarchical one is not a likely scenario in a random set of flux configurations. By the use of Random Matrix Theory we are in conditions to analyze this assertion.\\

%

\subsection{AdS Scale Separation}
Let us now classify  the scale separation between stable AdS vacua  $\Lambda_{\text{AdS}}$ 
and the squared mass corresponding to the lightest modulus for all models constructed from a Case A configuration. 
This study allows us to directly see, as shown in Figure \ref{fig:hierarTSU}, that by using a configuration of hierarchical fluxes it is more probable to find a hierarchy among moduli masses. Limited to our model we can say that the most probable scenario involves a maximum difference of masses of order of magnitude 3 
where the difference is given by
\eq{
\Delta m^2 = \text{max}\, m^2 - \text{min}\, m^2 \,.
}
Notice that an exponential $\Delta m^2$ as present in a KKLT model is discarded in our case, probably due to the fact that we are considering a hierarchy among fluxes of an order of magnitude between 1 and 4 
which in turn is a consequence of Bianchi and Tadpole constraints \cite{Betzler:2019kon}.\\

The AdS swampland scale conjecture asserts that it is not possible to separate the size of the AdS space and the mass of its lightest mode beyond a certain limit, this is
\eq{
\left( \text{min} \, m^2 \right) L^2_{\text{AdS}} \leq c \,,
}
where $c$ is constant of order 1, and  $L^2_{\text{AdS}} \sim \Lambda^{-1}_{\text{AdS}}$. This conjecture is motivated from the point of view of the KKLT scenario, in the sense that any uplifting mechanism (from a supersymmetric stable vacua) does not destabilize the K\"ahler moduli as far as the potential well is parametrically narrow in comparison with the 
energy gap that needs to be filled by the uplifting mechanism. For the KKLT scenario, indeed this criteria is not fulfilled and thus it raises the question of its validity \cite{Gautason:2018gln}. \\

We analyze this conjecture for our simple model (see Figure \ref{fig:AMSSC}) and we observe that the effect of  using a hierarchical flux configuration solution implies an increase on  
the probability to find a scenario in which the scale between the lightest modulus and the size of AdS space be in the lower bound. 
Thus as argued by \cite{Gautason:2018gln}, any attempt to uplift the AdS vacua may destabilize the lightest modulus. However, for fluxes without hierarchy (yellow bars) it is observed that the mean value of the vacua is around 1 saturating the bound. Therefore we conclude that a hierarchical flux configuration leads us to scenarios in which the ratio ${\rm min}\, m^2/\Lambda_{\rm AdS}<1$ 
, which according to the conjecture of AdS scales, would produce an instability if uplifting to dS. \\

In summary, \textit{by assuming a hierarchy on the flux configuration among different sectors (NS-NS, R-R and NG) it is more probable for the generated vacua to have small values for the vacuum energy and a small value for the string coupling. Also, scenarios constructed with non-hierarchical fluxes exhibit a higher probability for the lightest modulus to be much larger than the cosmological constant, which according to a recent conjecture, makes impossible to uplift to a dS vacuum.}\\

However, among all possible flux configurations, having a hierarchical one is not a likely scenario in a random set of flux configurations. By the use of Random Matrix Theory we are in conditions to analyze this assertion.\\

\subsection{Relation to Random Matrix Theory}
The refined swampland criterion implies that for a dS vacuum the lowest eigenvalue of the mass matrix shall be negative and thus unstable. Indeed, if the RdSC is not satisfied, there exist an instability which leads to a breakdown in entropic arguments \cite{Ooguri:2018wrx}. This line of thought leads us to consider some sort of information/probabilistic feature of the dS conjecture and its refinement. Within this context, it was found \cite{Low:2020kzy} that using random functions as scalar potentials, the dS conjecture as well as the refined dS conjecture are the result of the most probable scenario. However, the connection with real vacua coming from dimensional reduction in string theory was not clear. \\

\begin{figure}[!htbp]
   \centering
   \includegraphics[scale=0.29]{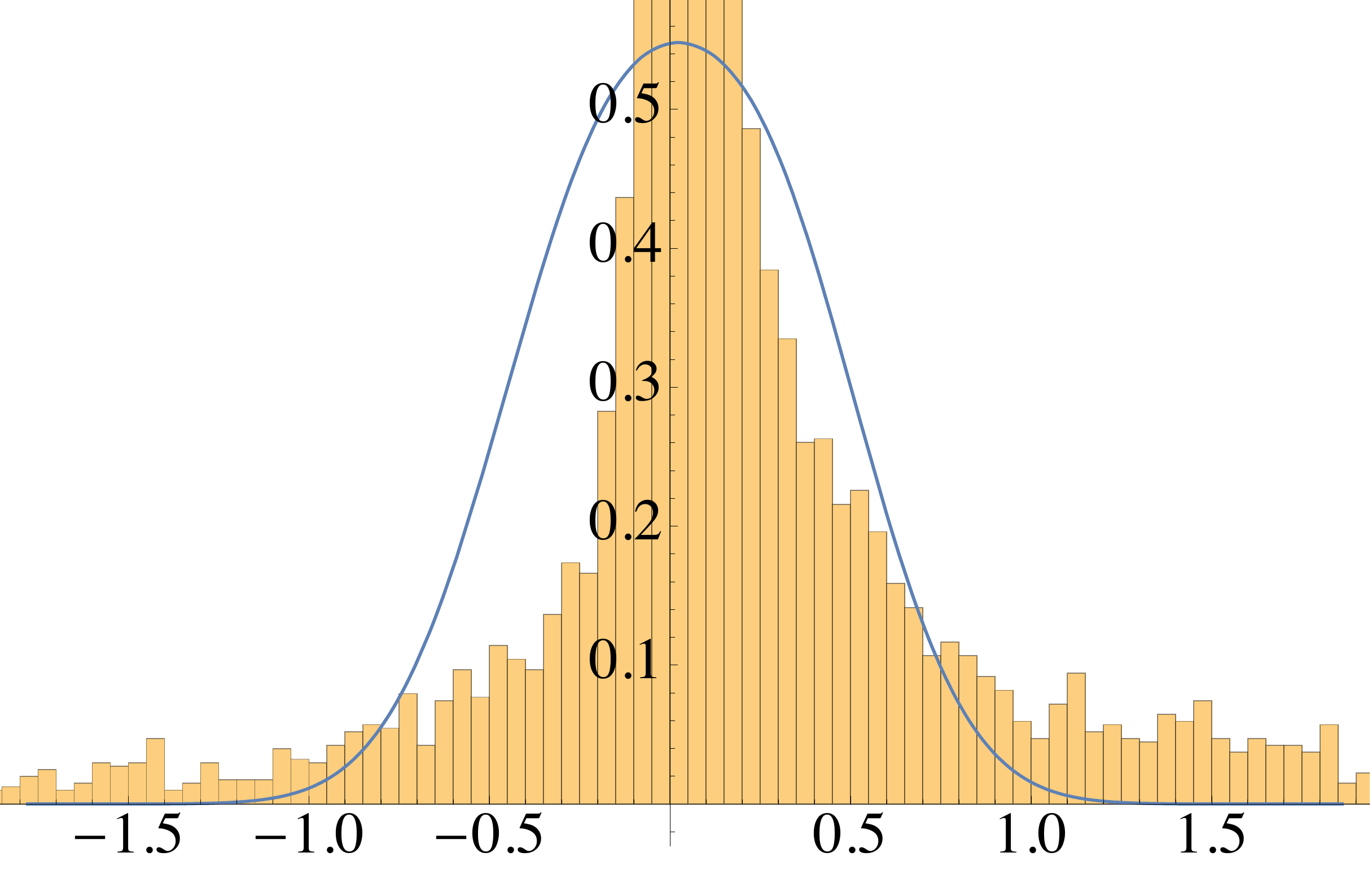} 
      \caption{\emph{Density distribution for all the eigenvalues of the mass matrix.}}
        \begin{picture}(0,0) 
   \put(80,55){$\text{Eig} \nabla_i \nabla_j V$}
   \put(-10,172){PDF}
  \end{picture}
   \label{fig:pdf-vacua}
\end{figure}
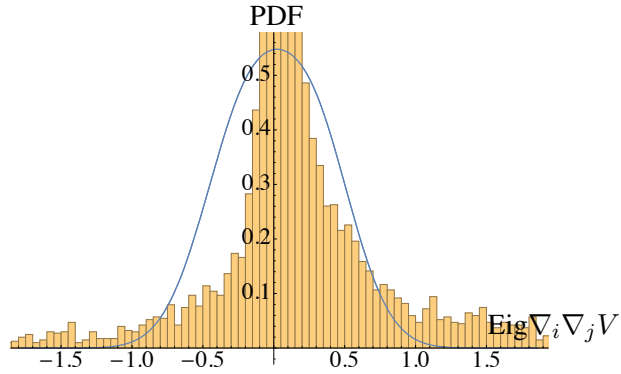

As already mentioned, after combining genetic algorithms and neural networks, we realize that there is a low probability of finding critical points.
In Figure \ref{fig:pdf-vacua} we present the histogram of the probability density distribution of the critical points obtained by all  flux configurations. This distribution presents a mean value  of 0  and a standard variation $\sigma=$0.35. Besides, assuming identical and independent distributed (i.i.d.) entries coming from a Gaussian distribution, the  probability density function (PDF) of the eigenvalue $\lambda$-spectrum of the mass matrix can be calculated by \cite{mehta2004random} (for a kindly check of the calculations see \cite{livan2018introduction})
\eq{\label{eq:pdf}
\rho (\lambda) = \frac{\mathcal{N}}{\sigma} \sum_{k=0}^{N/2-1} \exp\left[ \frac{\lambda^2}{2 \sigma^2} \right] \left( R_{2k} (\lambda) \Phi_{2k+1} (\lambda) - R_{2k+1} (\lambda) \Phi_{2k} (\lambda) \right) \,,
}
where $\mathcal{N} =\frac{N! |\hat a_N| 2^{N/2-1}}{N \mathcal{Z}}$, $\hat{a}$ is a constant that depends on $N$, $\mathcal{Z}$ is a normalization factor analogous to the partition function (see \cite{mehta2004random})  and $N$ is the rank of the mass matrix. The functions $\Phi_k(\lambda)$ are given by
\eq{
\Phi_k \left( \lambda \right) = \frac{1}{\sigma} \int_{-\infty}^{\infty} d\lambda' R_k \left( \lambda' \right) \exp\left[ \frac{\lambda'^2}{2 \sigma^2} \right] \text{sign} \left( \lambda - \lambda' \right) \,,
}
with $R_k$ being essentially Hermite polynomials:
\eq{
R_{2k} \left( \lambda \right) &= \frac{\sqrt{2}}{\pi^{1/4} 2^k (2k)!!} H_{2k} \left( \lambda \right) \,, \\
R_{2k+1} \left( \lambda \right) &=  \frac{\sqrt{2}}{\pi^{1/4} 2^k (2k)!!} \left[ -H_{2k+1} \left( \lambda \right)  + 4 k H_{2k+1} \left( \lambda \right) \right] \,. \\
}
Thus, although we do not know to which probability density distribution the entries of the mass matrix belong, we shall assume that a Gaussian distributions comes as  a good approximation, and it serves as a limiting case (see solid line of Figure \ref{fig:pdf-vacua} which represents the PDF given by Eq. \ref{eq:pdf}). We expect that a much amount of data would make closer our mass eigenvalues PDF comes from a GOE spectrum. Hence, the rest of our analysis relies on this assumption.\\

\begin{figure}[!htb]
\centering
\begin{tikzpicture}
\draw[thick,->] (0,0) -- (10,0);
\draw[thick,->] (5,-1) -- (5,4) node[left]{$\rho \left( \lambda, N \right) $};
\draw[fill,gray] (0,0) -- (1.0,0) to [out=1, in = 180] (2,2) to [out=0, in=181] (3.0,0) ;
\draw[dashed] (2,0) -- (2,4);
\draw[dashed,red] (4.5,0) node[above]{-c'V} -- (4.5,3.5) node [left]{AdS};
\draw[dashed,->,red] (4.5,1.65) -- (3.2,1.65) node[below]{Favored AdS};
\draw[dashed,green] (5.5,0) node[above]{c'V} -- (5.5,3.5) node [right]{dS};
\draw[dashed,->,green] (5.5,2.25) -- (6.7,2.25) node[above]{Unfavored dS};
\draw (1,1.6) node[below]{$N^{-1/6}$} (0.5,2.1) node[below]{Tracy-Widom};
\draw (3.0,0) node[above]{min $\nabla_i \p_j V$} (10.1,-0.7) node[above]{Eig ($\nabla_i \nabla_j V$)};
\draw[thick,blue] (7.7,0) node [below] {$\sqrt{2N}$} (8,0) arc (0:180:3cm)  node [below] {$-\sqrt{2N}$};

\end{tikzpicture}
\caption{\emph{ Statistics for the Extreme value statistics for the GOE. The blue line represents the probability density function for the eigenvalues.  The horizontal axis represent the eigenvalues of the mass matrix, whereas the vertical axis represents the probability density function of those eigenvalues. For the case of AdS, the RdS conjecture is interpreted as the  probability to find the minimum below zero, i.e., $\mathbb{P} \left( \text{min} \, \nabla_i \nabla_j V < c' V \right)$ (red lines) which is easily achieved. For the case of dS vacua, the RdS conjecture translates into the probability of all eigenvalues to be positive and above the $c' V$ vertical line, i.e., $\mathbb{P} \left( \text{min} \, \nabla_i \nabla_j V < c' V \right)$ which is in  general hard to be achieved (green lines).}}
\label{fig:twRM}
\end{figure}
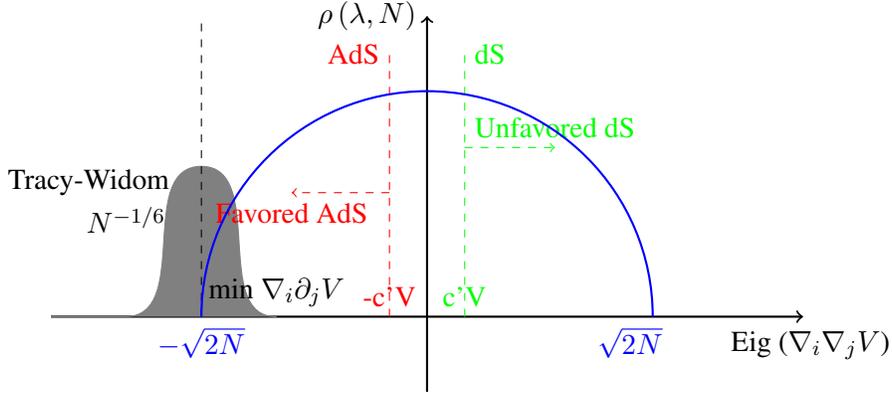

Now, 
if the mass matrix is interpreted as a random matrix with identically and independently distributed entries with Dyson index 1, this is a Gaussian Orthogonal Ensemble (GOE) with real entries, it is quite unlikely to get only positive eigenvalues. This well known result from random matrix theory (RMT) follows from the fact that extreme eigenvalues of a GOE obey the Tracy-Widow statistics and that any fluctuation in the lower limit is suppressed by a power $N^{-1/6}$ for $N$ be the rank of the matrix \cite{dean2006large}  (as shown in Figure \ref{fig:twRM}). Thus let us put the RdSC in terms of a RMT.\\

The eigenvalues of a random matrix are expected to be distributed around zero, however, for large $N$ it has been proved that the minimum eigenvalue tends to $-\sqrt{2N}$ while the maximum to $\sqrt{2N}$.  As we said, fluctuations of extreme eigenvalues falls as $N^{-1/6}$, and thus allowing a possibility for the minimum eigenvalue to acquire a value different from $-\sqrt{2N}$. The distribution of fluctuation around $-\sqrt{2N}$ is shown by the shadow region in Figure \ref{fig:twRM}. For a large value of $N$ it seems that
\eq{
(\text{min} \nabla_i \nabla_j V )_{RM} \leq \alpha,
}
where the subindex $RM$ stands for a random mass matrix and $\alpha$ a number to be determined. It is expected that in such scenarios (eigenvalue probability distribution),  the probability for the minimum eigenvalue to be negative increases as $N$ increases. Actually, as proved in \cite{dean2006large}, the probability for the minimum eigenvalue to be bounded by a number $t$ is given by
\eq{
\mathbb{P} \left( \text{min} \, \lambda > t \right) = \exp \left[ -\frac{1}{24}  \left| \sqrt{2} N^{1/6} (t+ \sqrt{2 N} )\right|^3 \right] \,.
}
Notice that for $t >0$($<0$) $\mathbb{P}$ reduces (increases). In our case in which the eigenvalues $\lambda$ are related to the mass eigenvalues, i.e. $\lambda\rightarrow  \text{Eig}\, ( \nabla_i\nabla_j V )$ we can chose $t$ to be the proportional to the potential at the minimum. In that case we see that for $N=6$,
\eq{
\mathbb{P} \left( \text{min} \, \nabla_i \nabla_j V > c' V \right) = \exp \left[ -\frac{1}{24}  \left| \sqrt{2} \cdot 6^{1/6} (c'V + 2\sqrt{3} )\right|^3 \right] \,,
}


Thus for a dS vacua, $\mathbb{P}$ is very small and the  largest the value for $V$ at the minimum, the smaller the probability for  the lightest moduli to be positive. dS vacua seem to be very less favored than unstable critical dS points. Similarly, for  an  AdS vacuum,  the probability for having all positive eigenvalues is much higher than the corresponding for a dS extreme point and it raises as the absolute value of the vacuum energy grows (see Figure 
\ref{fig:twRM}). We then conclude that the most probable configurations satisfy the bound
\eq{
\text{min} \nabla_i\nabla_j V\leq -c' V, 
}
in agreement with the RdSC.\\

Notice as well that the probability expression also asserts that the ratio between the minimum squared mass in a stable AdS vacuum and the AdS scale larger than one, this is,  $\text{min } m^2/\Lambda_{AdS}<1$ is more favored. Hence the AdS scale conjecture is also encoded in this probabilistic interpretation.
%
Taking all our observations together, we conclude that:\\

\hspace{0.5cm} \emph{In an effective model constructed from  a perturbative flux compactification (at least for an isotropic toroidal one) the probability for the minimum mass eigenvalue to be larger than  the corresponding vacuum energy $\Lambda$ is given by 
\eq{
\exp \left[ -\frac{1}{24}  \left| \sqrt{2} \cdot 6^{1/6} (\Lambda+ \sqrt{12} )\right|^3 \right]. \nonumber}
This implies that the most probable mass configurations with positive value of the cosmological constant are those which  contain negative mass states in its spectrum. For the case of a negative value of the negative cosmological constant, the most probable scenario implies the presence of Tachyons. Notice that this implies that the most probable effective models are those precisely satisfying the RdS and the AdS conjectures.} 

\section{Final comments}
\label{sec:cuatro}
In this work we have implemented a vacuum search through an Artificial Neural Network coupled to a Genetic Algorithm. We report more than 60.000 flux configurations yielding to a scalar potential with at least one critical point. 
We use a simple model consisting on type IIB string theory flux compactification on an isotropic torus including non-geometric fluxes.  With the data obtained by this classification we can test $-$in terms of probabilities$-$ some of our model's features in the light of recent Swampland conjectures.\\ 

Our main conclusion is that, at least for the studied model, generic flux configurations produce different 
vacua with two clear features:
\begin{itemize}
\item
The Refined dS Conjecture is fulfilled and the relation $\text{min}\, \nabla_i\nabla_jV\leq -c' V$ with $c'$ of order 1 is graphically proved in Figure \ref{fig:RdSC}. Notice the absence of certain stable AdS  as well as some unstable dS vacua.
\item A statistical correlation is observed favoring a small value for the cosmological constant in models exhibiting a small string coupling. 
\end{itemize}
Our results show a clear increase in probability to find vacua with a smaller than unit cosmological constant (and in consequence within the perturbative regime) if they are constructed from a hierarchical flux configuration, 
meaning a flux configuration in which the integer quantized values for the different sectors, including non-geometrical fluxes, differ by at least one order of magnitude. The construction of different vacua, stable or not, from a hierarchical flux compactification leads to the following facts:
\begin{itemize}
\item
The value of the corresponding cosmological constant is small and in consequence within the range of a perturbative effective theory. The probability to obtain such vacua  increases by selecting the RR sector with the highest flux values, which in turn makes the Complex Structure moduli the heaviest.
\item
The probability to have an AdS stable vacuum in which the lightest modulus is heavier than the corresponding cosmological constant increases. Those scenarios seem to exhibit a persistent difficulty to be uplifted to a dS vacuum.\\
\end{itemize}


We also observe by the use of Random Matrix Theory that stable vacua are much less probable than unstable ones. Actually, in a random selection of fluxes which present a Gaussian distribution of mass eigenvalues, the more probable vacuum solutions are those which precisely fulfill the Swampland conjectures, namely the Refined de Sitter and the Ads scale ones. This suggests that the origin of the Swampland constraints, at least for the models we have studied, is probabilistic.\\

Finally we notice that the possibility to select a hierarchical flux configuration from a random set of different flux configurations, is very low, indicating that for a hierarchical flux configuration to be the source of effective models, 
a high-energy process must be the cause of fixing the values for fluxes. We leave this important issue for a future work.\\

\begin{center}
{\bf Acknowledgments}
\end{center}
We thank Alejandro Cabo, Anamar\'{\i}a Font, Vishnu Jejjala, Albrecht Klemm, Challenger Mishra and Fernando Quevedo for useful discussions and comments. We thank Miguel Sabido for kind support  through the Data Lab at University of Guanajuato. N.C.B. and O.L.B. are supported by the project CIIC 290/2020 UGTO, CONACyT Project A1-S- 37752 and CONACyT Project CB-2015-01-258982 . C.D. is supported by CONACyT through the S.N.I. program, and D.K.M.P. by DAIP Universidad de Guanajuato under project CIIC 344/2019 and by the Simons Foundation Mathematical and Physical Sciences Targeted Grants to Institutes, Award ID:509116. J.A.M.B. thanks the National Council of Science and Technology (CONACyT), Mexico, for his Assistantship No. CVU- 736083.

\appendix

\section{Artificial Neural Network}
\label{sec:app3}
Artificial neural networks (ANNs) are algorithms inspired in the biological learning process. The use of machine learning techniques to solve classification problems has attracted more attention in the last years \cite{Bao:2020sqg,Halverson:2020opj,Parr:2019bta,Cole:2019enn,Brodie:2019dfx,He:2019vsj,He:2018jtw}. The machine learning techniques allows to search in large amount of data for specific patterns and thus, it provides an exhaustive check in a short time\footnote{For instance, in \cite{Parr:2019bta} using data mining the authors are able to look for suitable heterotic compactifications that selects an appropriate line bundle to produces a phenomenological motivated extension of the standard model.}. \\

In the following we describe in simple terms, the structure of an ANN.
 Each neuron in the hidden layer is connected with all the neurons in the neighboring clusters through a weight factor. The weighted interconnection is quantified by
\begin{equation}
a^l_j = \sigma \left( \sum_{k} w_{jk}^l a_k^{l-1} + b_j^l \right) \,,
\end{equation} 
where the sum is over all neurons in the $k$ level in the $l-1$ layer, the weights $w^l$ connects each $l$-th layer of neurons, $b^l$ is the bias factor and $a^l_j$ is the response/entrance of the ANN, for instance at $l=1$, $a$ represents the input data and at $l=n$ (last layer) $a$ represents the output of the ANN. The function $\sigma( \cdot )$ is the activation function which is a sigmoid function that introduces the non-linearities to the ANN. The weights and bias factors are determined in such a way that the mean relative error defined as
\begin{equation}
\text{MSE} = \frac{1}{n} \sum_{i=1}^n \left( a_n - t_n \right)^2 \,,
\label{eq:mse}
\end{equation}
is diminished, where $a_n$ is the response of the ANN an the $t$ is the target value. The numerical values are determined using the Levenberg-Marquardt optimization, which is a deterministic algorithm for non-linear systems that is able to find local minima in a iterative manner. This optimization requires to minimize the regularized function
\begin{equation}
f(w) = \frac{1}{2} (r_i)^2 + \frac{1}{2}\gamma^k (w^{k+1}_i-w^k_i)^2 \,,
\end{equation}
where $r_i =a_i(w)-t_i$ is the residual and $\gamma^k$ is a regularization parameter which is chosen through the trust region approach \cite{suratgar2005modified}. Thus, at each step the weights are determined by solving the equation
\eq{
\left( J_{il}(w^k) J_{lj} (w^k) -\gamma^k \delta_{ij} \right) \left( w^{k+1}_j-w^k_j \right) = - J_{il}(w^k) r_l (w^k) \,,
}
for $w^{k+1}$ at each iteration, where the index $k$ represents the $k$-iteration and $J_{ij} = \partial_i r_j$ is the Jacobian of the residual.\\


\subsection{An example of an  ANN}
\begin{figure}[!htbp]
   \centering
   \includegraphics[scale=0.8]{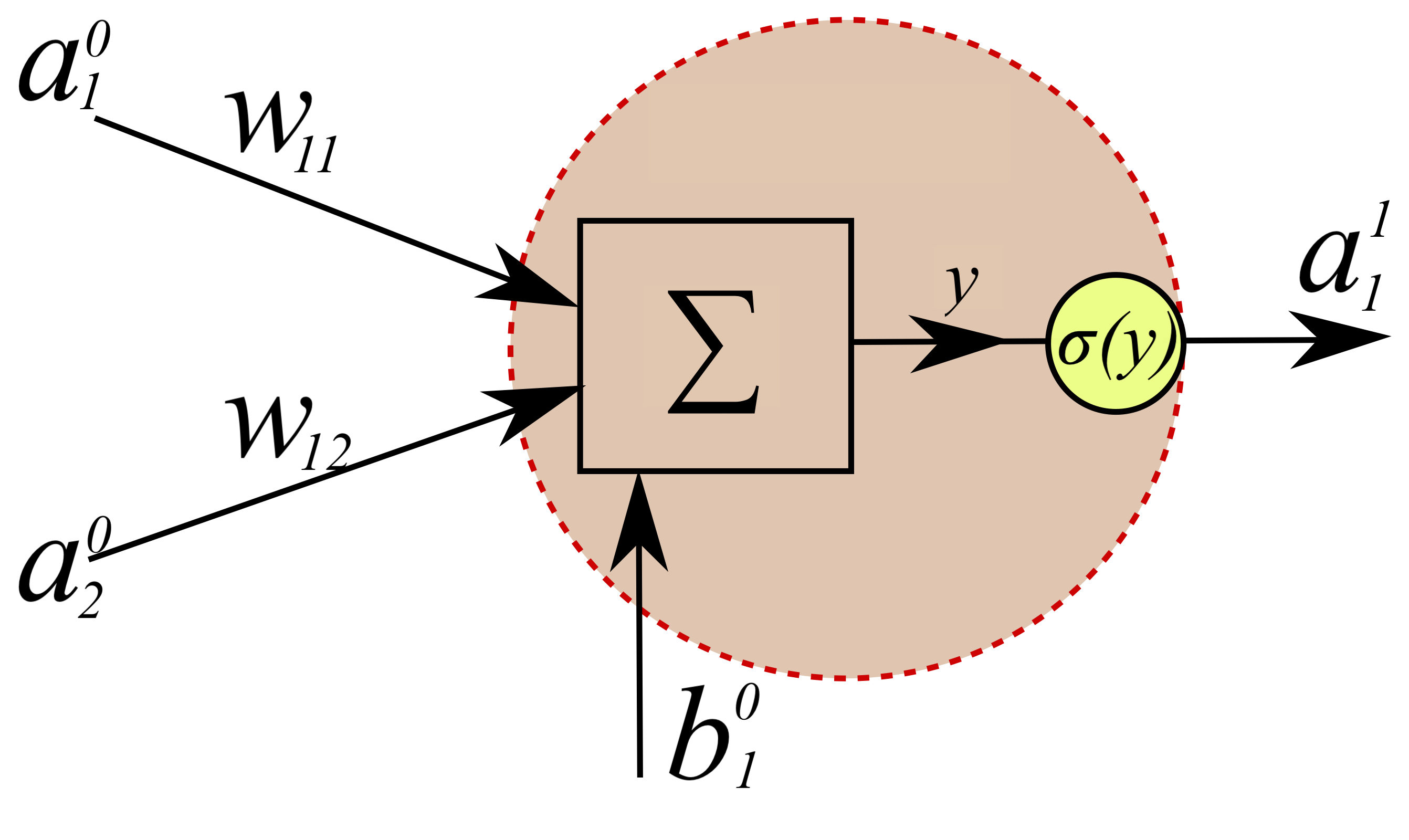} 
      \caption{\label{fig:AND}{\it  Representation of a single Neuron. For this example the input is described by the vector $(a^0_1,a^0_2)$, associated to the neuron
      are the weights $w_{11},w_{12}$ and the bias $b^0_1$, with $y=w^T.a^0+b^0_1$. With the neuron output  $a^1_1=\sigma(y)$ i.e. the activation function
      evaluated at $y$.}}
\end{figure}

To clarify the algorithm to compute the output on a ANN, in this section we present an explicit example of how a single perceptron is trained in order to reproduce the bolean product operation. Recall that this binary operator requires as input data two bits and calculates an output binary data (see Figure \ref{fig:AND}). The training set employ the for possibilities for the bolean product, namely, $0 \cdot 0 = 0$, $0 \cdot 1 = 0$, $1 \cdot 0 = 0$ and $1 \cdot 1 = 1$. \\ 

\begin{figure}[!htbp]
   \centering
   \includegraphics[scale=0.5]{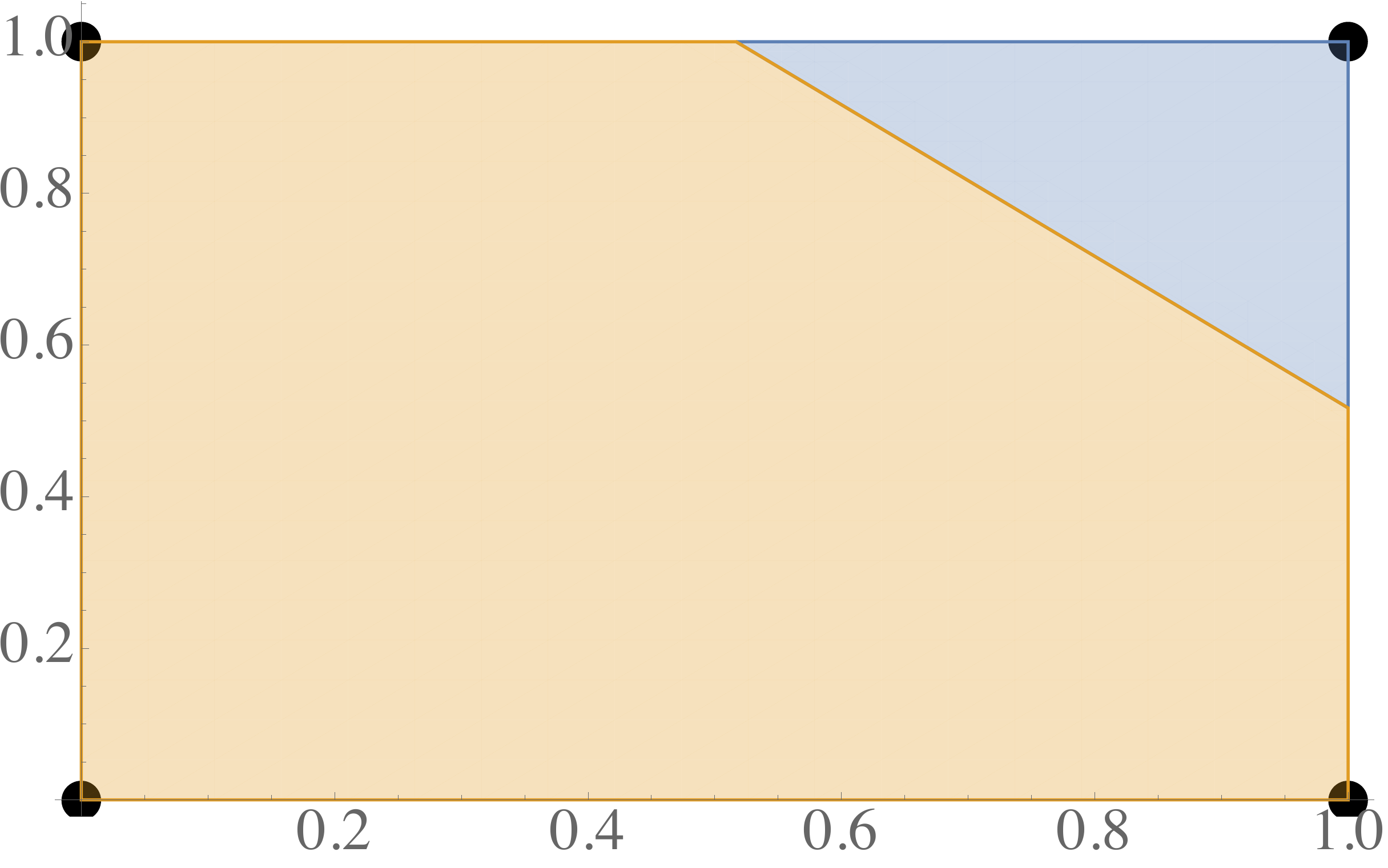} 
       \begin{picture}(0,0) 
   \put(-180,-5){First bit}
   \put(-370,100){\rotatebox{90}{Second bit}}
  \end{picture}
      \caption{\label{fig:Clas} Solution for the case of the AND gate. The x axis represents the first bit  of the AND gate whereas the y axis the second bit. Thus, the yellow regions defines the region where the bolean product is 0 whereas the blue region represents the region where the bolean product is 1. The sharp changes between the region 0 and 1 is a consequence of the neuron activation.}
\end{figure}
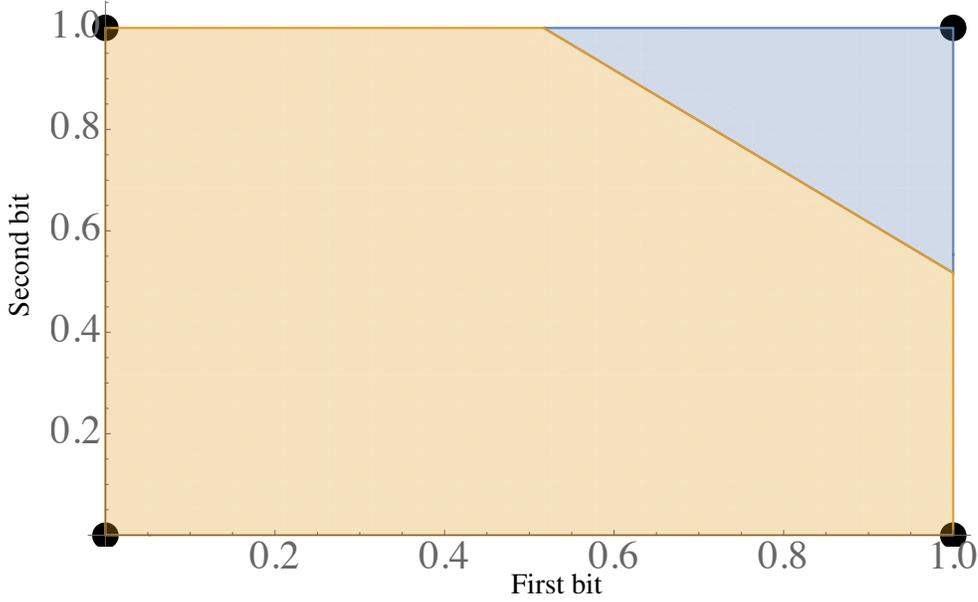

Thus, for the single perceptron case, we initialize the weights of the network, for instance $w_{11}^0 = 0.5$, $w_{12}^0 = 1.5$ and for the bias $b_1^0 = 3$. For concreteness let us consider as activation function the Logistic Sigmoid Function
\eq{
\sigma ( y ) = \frac{1}{1+\text{exp} \left( -y \right) } \,, 
}
and for the first case let us consider the bolean product ($0 \cdot 0 = 0$), the argument of the activation function is
\eq{
y_i = \sum_{j} w_{ij}^1 a_j^0 = 3
}
and thus the activation function  is $\sigma (3)  = 0.95$. This result tell us that for the selected weight values the neuron gets activated with an answer of 1. However, the correct answer shall be 0 (since we are evaluating the $0 \cdot 0 = 0$ case, and we have obtained high error for the first bolean operation. Now, computing the remaining cases we get
\eq{
y_0 = w_{01}^1 a_1^0 + w_{02}^1 a_1^0 +b_0  &= 0.5 \cdot 0 +1.5 \cdot 0 + 3 = 3.00 \rightarrow \sigma (3.00) = 0.95 \\
y_0 = w_{01}^1 a_1^0 + w_{02}^1 a_1^0 +b_0 &= 0.5 \cdot 0 +1.5 \cdot 1 + 3 = 4.50 \rightarrow \sigma (4.50) = 0.98 \\
y_0 = w_{01}^1 a_1^0 + w_{02}^1 a_1^0 +b_0 &= 0.5 \cdot 1 +1.5 \cdot 0 + 3 = 3.55 \rightarrow \sigma (3.55) = 0.97\\
y_0 = w_{01}^1 a_1^0 + w_{02}^1 a_1^0+b_0  &= 0.5 \cdot 1 +1.5 \cdot 1 + 3 = 5.05\rightarrow \sigma (5.05) = 0.99 \\ 
}
geometrically it is possible to see that in a $a_0^0 \,\text{vs} \, a_1^0$ plot, the region is divided into two regions, namely, if the value of the function $y_j>0$ the neuron is activated and for $y_j<0$ the neuron is de-activated as it is shown in Fig. \ref{fig:Clas}. Thus, for the selected weights all the neuron is activated in all the training cases and the MSE given by Eq.\ref{eq:mse}   is 0.698, which is too high to be acceptable. This step know as forward propagation allow us to compute the error for the randomly selected weights. The next step is to modify the random selected weights in order to reduce the MSE. The simplest way to do it is by implementing the batch gradient descent algorithm, which calculates the gradient of maximum descent for some objective function. Thus, we want to know how the gradient changes the objective function as a the weights changes, this is done by applying the rule of chain as

\eq{
\frac{\p \text{MSE}}{\p w_{0j}} = \frac{\p y_j}{\p w_{0j}} \cdot  \frac{\p f}{\p y_j} \cdot \frac{\p \text{MSE}}{\p f} \,, \\
}
where we shall use $w_{03} = b_0$. Thus, the new value of the weights, can be computed by resting the gradient from the previous value
\eq{
w_j = w_j - \alpha \frac{\p \text{MSE}}{\p w_{0j}} \,,
}
where $\alpha$ is known as the learning rate and is an parameter between 0 and 1 (for this case we shall take 1). Thus, after the first iteration, this algorithm allows us to update the weights as,  $w_{01}^0 = 0.48$, $w_{02}^0 =1.48$ and for the bias $b_1^0 = 2.95$ with a MSE of $0.695$. This algorithm usually has a low convergence ratio, since it takes a lot number of steps to achieve a desired convergence criteria however for this toy model example a MSE $< $0.002 is achieved at 2000 iterations. Thus, once the desired MSE is obtained, the updated weights take the values $w_{01}^0 =5.49$, $w_{02}^0 =6.45$ and the bias $b_1^0 = -8.28$ with an MSE of $0.002$. Thus, a forward calculation shall make the ANN to reproduce the bolean product operation as,
\eq{
y_0 = w_{01}^1 a_1^0 + w_{02}^1 a_1^0 +b_0  &= 5.49 \cdot 0 +6.45 \cdot 0  -8.28 = -8.28 \rightarrow \sigma (-8.28) = 0.00, \\
y_0 = w_{01}^1 a_1^0 + w_{02}^1 a_1^0 +b_0 &= 5.49 \cdot 0 +6.45 \cdot 1  -8.28 = -1.83 \rightarrow \sigma (-1.83) = 0.05,\\
y_0 = w_{01}^1 a_1^0 + w_{02}^1 a_1^0 +b_0 &= 5.49 \cdot 1 +6.45 \cdot 0  -8.28 = -2.79 \rightarrow \sigma (-2.79) = 0.05,\\
y_0 = w_{01}^1 a_1^0 + w_{02}^1 a_1^0+b_0  &= 5.49 \cdot 1 +6.45 \cdot 1  -8.28 = 3.66\rightarrow \sigma (3.66) = 0.93. \\ 
}
which approximately is the desired result. Notice that for this toy model we employ all the available data to train the ANN and there are no remaining data for the validation. However, for a more complicated training set, a subset is not used in the training 
and a validation set  is used. 

\section{Isotropic Toroidal Compatifications with Non-Geometric Fluxes}
\label{sec:app:torus}

In this appendix we describe a type IIB string theory compactification with non geometric fluxes on
an isotropic $T^6$ torus. 
The effective 4D $\mathcal{N}=1$ theory can be obtained in terms of a superpotential given by
\begin{equation}
W=P_1(U)-iSP_2(U)+iTP_3(U),
\end{equation}
with $U,S$ and $T$ being the complex structure, axio-dilaton and K\"ahler moduli respectively.  $P_1(U)$, $P_2(U)$
and $P_3(U)$ are polynomia depending on $U$ and their coefficients are given in terms of
NS-NS $h_j$, R-R $f_i$ and non geometric $b_k$ fluxes respectively. The polynomia depend on
the fluxes as follows
\begin{eqnarray}
P_1&=& f_1 + 3 I f_2U - 3f_3U^2 - I f_4U^3, \\
P_2&=&h_1 + 3 I h_2 U - 3h_3U^2 - I h_4U^3 , \nonumber  \\
P_2&=& 3 b_1+I (2b_2+b_3)U-(2 b_4+b_5) U^2-I b_6 U^3. \nonumber\\
\end{eqnarray}
This structure of the superpotential, where the fluxes $h$ and $b$ determine the relevance of $S, T$ to the scalar potential dependence respectively,
suggests that the hierarchy between moduli masses can be reached
by implementing hierarchies among the fluxes.
The K\"ahler potential reads
\begin{equation}
K=-\ln (S+S^\ast)-3\ln (U+U^\ast)-3\ln (T+T^\ast).
\end{equation}
We decompose the scalar fields in terms of their real and
imaginary components as: $U=u+iv$, $S=s+ic$ and $T=t+i\tau$,
where $u,v,s,c,t,\tau$ are real fields.\\

The corresponding scalar potential can be computed in terms of $K$ and $W$ and
is given by
\begin{equation}
V=e^K\left( |D_IW|^2K^{IJ}-3|W|^2\right).
\end{equation}
This potential has extremal values when SUSY is preserved in an AdS or Minkowski vacuum. When SUSY is not preserved, it is possible to have extrema for diferente values of $V_0$.  
The appearance of these extrema follows from the presence of fluxes including non-geometric fluxes, which could stabilize the K\"ahler moduli $T$. However, different flux configurations  produce different type of vacua whose characteristics are also constrained by  the tadpole on the NS-NS and RR fluxes. One important aspect in the construction of vacua solutions is to obtain physical consistency. This implies having a positive larger than one value for $s=1/g_s$. This
requirement ensures that the perturbative approximation for IIB string compactification on the isotropic torus is valid. \\

Since we search for SUSY vacua, it follows that all K\"ahler derivatives must vanish, i.e.
\begin{equation}
D_U W=D_S W= D_T W=0.
\end{equation}

\subsection{Small cosmological constant: a toy example}

We present an analysis concerning some SUSY solutions in order to elucidate the possible existence of some generic conditions on the flux configuration which leads us to effective models in which for an  AdS vacuum not only $s>1$ but also $|V_0| \ll 1$. Our goal is to obtain some insights about the characteristics of different flux configurations, which can assure the construction of such desirable vacua.\\

In the following we take a particular path in  order to construct a supersymmetric solution with all moduli stabilized. First of all we observe that $D_TW=0$ implies that K\"ahler moduli are fixed to
\begin{equation}
t=\frac{3}{2} \qquad \text{and}\qquad \tau=-\frac{1}{2}\frac{\widetilde{p_3}}{p_3}=\frac{1}{2}
\end{equation}
where $P_2, P_3 \ne 0$ is assumed at any point of the complex structure at which the polynomial are evaluated with $p_3=-\widetilde{p_3}$. We are then forced to find some value $U_0$ for which this is valid for non-trivial polynomial. We  shall come back to this point. Meanwhile, $D_SW=0$ implies, once we take the above constraint fixed by $D_TW=0$, 
\begin{equation}
c=\frac{1}{2}~\frac{p_2+\widetilde{p_2}}{p_2-\widetilde{p}_2},
\end{equation}
while $s$ is kept unfixed. We shall fix it by minimizing the scalar potential with respect to $s$. Finally, $D_UW=0$ fixes $U=U_0$ as a solution of the equation\footnote{More constraints would involve a sharing solution with a similar equation for $P_1(U)$. See Ref \cite{CaboBizet:2019sku}.}
\begin{equation}
2u ~ \frac{\partial {P}_2}{\partial U} -3 P_2=0.
\end{equation}
However, as shown in \cite{CaboBizet:2019sku}, roots $U_0$ for the above polynomial implies that $P_3(U_0)=0$. In order to keep a SUSY solution we shall then assume that
\begin{equation}
\lim_{U\rightarrow U_0} \frac{\widetilde{p}_3(u_0,v_0)}{p_3(u_0,v_0)}=-1
\end{equation}
with $\tau(u_0,v_0)=1/2$. In this case the potential has the form
\begin{equation}
V(s)=-\frac{3}{2^7 u_0^3 t_0^3}\left(2 \text{Im}(P_2 \omega^\ast)+s|P_2|^2+\frac{|\omega|^2}{s}\right),
\end{equation}
where $\omega (u_0,v_0)= (p_1+c_0 p_2)+i(\widetilde{p}_1+c_0\widetilde{p}_2)$. It follows that the string cuopling is fixed at the value
\begin{equation}
s^2_0=\frac{|\omega|^2}{|P_2|^2}(u_0,v_0),
\end{equation}
at the minimum of the scalar potential. At this point, our interest focuses in finding general conditions on the flux configuration upon which  $|V_0|<1$ while $s_0>1$. For the latter, observe that an easy and direct way to assure a small string coupling is to have $|\omega|>|P_2|$ which can easily be obtained by considering a hierarchy on the flux configuration among RR, NS-NS and non-geometric fluxes. Since these fluxes enter as the real coefficients on the polynomial $P_i(U_0)$ it follows that if RR fluxes are larger than NS-NS, which in turn are larger than non-geometric fluxes, one can obtain that $|\omega|>|P_2|$ for $\omega$ and $P_2$ evaluated at $U_0$.\\

To obtain a small value for $V$ at the minimum, observe that
\begin{equation}
V_0=-\frac{1}{2^3 u_0^3 3^2}( \text{Im}(P_2 \omega^\ast)+|\omega||P_2|),
\end{equation}
which by taking same order fluxes among the same type of fluxes (e.g., all RR fluxes are the same order but larger than all NS-NS, which are of the same order among them), one can assure that $\text{Im}(P_2(U_0)\omega^\ast(U_0))$ be smaller than unit. Hence we can expect a small value for $|V_{min}|$ if $\omega$ and $|P_2|$ at $U=U_0$ are also smaller than one. As shown in \cite{CaboBizet:2019sku},  $\omega \sim {\cal O}(1)$ in flux units. This implies that 
\begin{equation}
V_0\sim -\frac{{\cal O}(1)}{u^3_0 s_0}.
\end{equation}
In effect, having a hierarchy on different type of fluxes, one can obtain in a generic form, at least for the SUSY solutions we have considered, a perturbative effective theory with a very small negative cosmological constant for $u_0>1/s_0$.\\ 

\bibliographystyle{JHEP}
\bibliography{references2}

\end{document}